\documentclass[twocolumn,showpacs,pre,amsmath,amssymb]{revtex4-1}

\usepackage{graphicx}
\usepackage{dcolumn}
\usepackage{bm}
\usepackage{color}

\def\b{\mathbf}

\begin{document}


\title{Lattice Mechanics of Origami Tessellations}

\author{Arthur A.~Evans$^1$}
\author{Jesse L. Silverberg$^2$}
\author{Christian D. Santangelo$^1$}

\affiliation{$^1$ Department of Physics, UMass Amherst, Amherst MA 01003, USA}
\affiliation{$^2$ Department of Physics, Cornell University, Ithaca, NY 14853, USA}


\begin{abstract}
Origami-based design holds promise for developing materials whose mechanical properties are tuned by crease patterns introduced to thin sheets. Although there has been heuristic developments in constructing patterns with desirable qualities, the bridge between origami and physics has yet to be fully developed. To truly consider origami structures as a class of materials, methods akin to solid mechanics need to be developed to understand their long-wavelength behavior. We introduce here a lattice theory for examining the mechanics of origami tessellations in terms of the topology of their crease pattern and the relationship between the folds at each vertex. This formulation provides a general method for associating mechanical properties with periodic folded structures, and allows for a concrete connection between more conventional materials and the mechanical metamaterials constructed using origami-based design.
\end{abstract}

\maketitle

While for hundreds of years origami has existed as an artistic endeavor, recent decades have seen the application of folding thin materials to the fields of architecture, engineering, and material science 
\cite{tachi2009generalization,tachi2010geometric,tachi2010one,hawkes2010programmable,dias2012geometric,schenk2013geometry,silverberg2014using}. Controlled actuation of thin materials via patterned folds has led to a variety of self-assembly strategies in polymer gels \cite{na2014programming} and shape-memory materials \cite{hawkes2010programmable}, as well elastocapillary self-assembly \cite{py2007capillary}, leading to the design of a new category of shape-transformable materials inspired by origami design. The origami repertoire itself, buoyed by advances in the mathematics of folding and the burgeoning field of computational geometry \cite{solomon2012flexible}, is no longer limited to designs of animals and children's toys that dominate the art in popular consciousness, but now includes tessellations, corrugations, and other non-representational structures whose mechanical properties are of interest from a scientific perspective. These properties originate from the confluence of geometry and mechanical constraints that are an intrinsic part of origami, and ultimately allow for the construction of mechanical meta-materials using origami-based design
\cite{tachi2009generalization,tachi2010geometric,tachi2010one,hawkes2010programmable,schenk2011folded,schenk2011origami,schenk2013geometry,wei2013geometric}. In this paper we formulate a general theory for periodic lattices of folds in thin materials, and combine the language of traditional lattice solid mechanics with the geometric theory underlying origami.


A distinct characteristic of all thin materials is that geometric constraints dominate the mechanical response of the structure. Because of this strong coupling between shape and mechanics, it is far more likely for a thin sheet to deform by bending without stretching. Strategically weakening a material with a crease or fold, and thus lowering the energetic cost of stretching, allows complex deformations and re-ordering of the material for negligible elastic energy cost. This vanishing energy cost, especially combined with increased control over micro- and nanoscopic material systems, indicates the great promise for structures whose characteristics depend primarily on geometry, rather than material composition. 

By patterning creases, hinges, or folds into an otherwise flat sheet (be it composed of paper, metal or polymer gel), the bulk material is imbued with an effective mechanical response. In contrast to conventional composites engineering, wherein methods generally rely on designing response based on the interaction between the constituent parts that compose the material, origami-based design injects novelty at the ``atomic" level; even single vertices of origami behave as engineering mechanisms \cite{abdul2013two}, providing novel functionality such as complicated bistability \cite{waitukaitis2015origami,hanna2014waterbomb,bende2014geometrically} and auxetic behavior \cite{schenk2011folded,schenk2011origami,schenk2013geometry,wei2013geometric}. This generic property inspires the identification of origami tessellations with mechanical metamaterials, or a composite whose effective properties arise from the structure of the unit cell. Although originally introduced to guide electromagnetic waves\cite{pendry2006controlling}, rationally designed mechanical metamaterials have since been developed that control wave propagation in acoustic media \cite{kadic2012practicability,seismic2014}, thin elastic sheets and curved shells\cite{farhat2009ultrabroadband,stenger2012experiments,shim2013harnessing,evans2013reflection}, and harness elastic instabilities to generate auxetic behavior \cite{zhang2008one,matsumoto2009elastic,bertoldi2010negative,matsumoto2012patterns,overvelde2012compaction}.
 
Traditional metamaterials invoke the theory of linear response in wave systems, but currently there is no general theory for predicting the properties of origami-inspired designs on the basis of symmetry and structure. In the following we propose a general framework for analyzing the kinematics and mechanics of an origami tessellation as a crystalline material. By treating a periodic crease pattern, we naturally connect the geometric mathematics of origami to the more conventional analysis of elasticity in solid state lattice structures. In section I we outline the general formalism required to find the kinematic solutions for a single origami vertex. In section II we discuss the general formulation for a periodic lattice, including both the kinematics of deformation modes and energetics for a periodic crease pattern. In section III we examine the well-known case study of the Miura-ori pattern. Our analysis here recovers known aspects of the Miura-ori pattern as well as identifies key features that have not been quantitatively discussed previously.

\begin{figure*}[!]
\includegraphics[width=.85\textwidth]{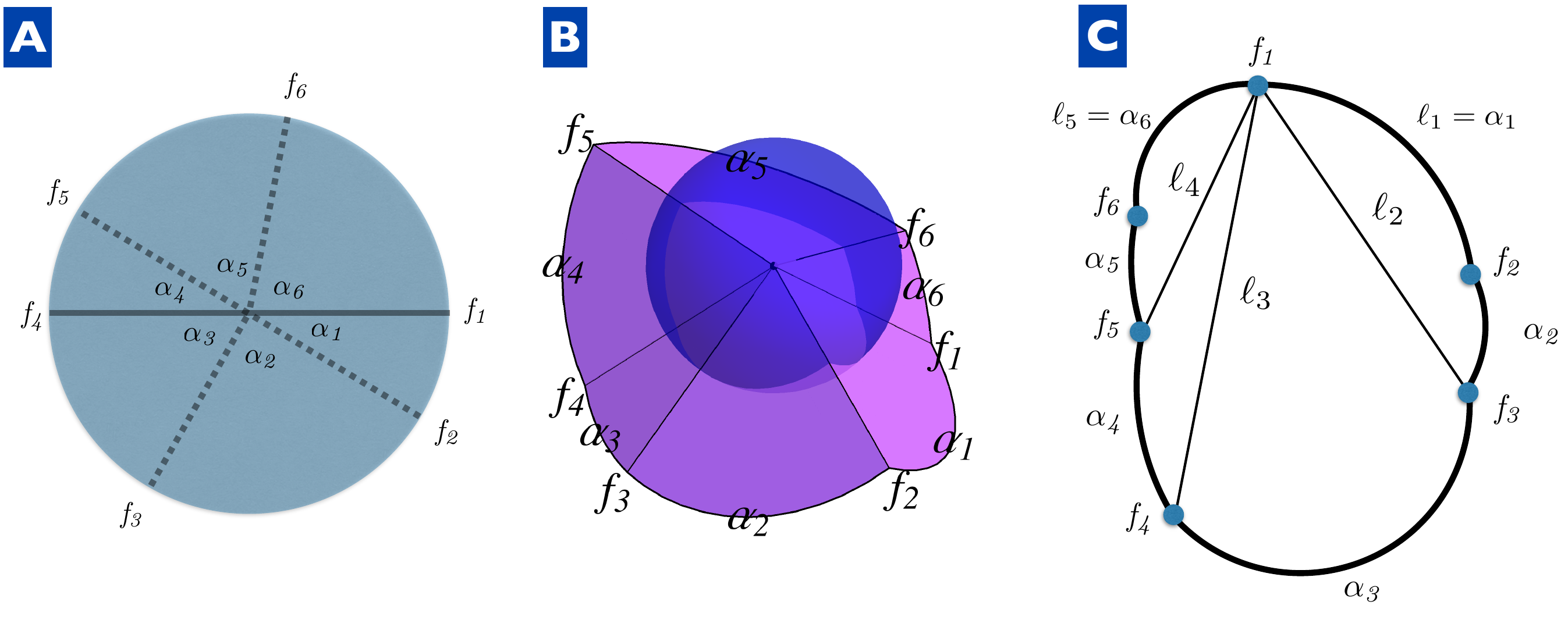}
\caption{\label{vertex}(color online) (A) Graph for a single vertex. This degree six vertex has its graph determined by the six sector angles $\alpha_i$. Each crease has a dihedral angle $f_i$ associated with it. In the flat case every $f_i=\pi$, or equivalently, every fold angle is identically zero, since the fold angle is defined as the supplement of the dihedral angle. (B) By assigning fold angles to each crease, a 3D embedding of the vertex (i.e. the folded form of the origami) is fully determined. Every face must rotate rigidly about the defined creases, and the sector angles must remain constant. There is a limited set of fold angles that will solve these conditions. (C) Schematic projection of the curve of intersection between the unit sphere and the folded form origami. For an $N$-degree vertex this projection generates a spherical $N$-gon. To proceed, the $N$-gon is divided into $N$-2 spherical triangles and the interior angles (i.e. the $f_i$) follow as a result of applying the rules of spherical trigonometry. All three dimensional origami structures are visualized using Tessellatica, a freely available online package for Mathematica \cite{tessellatica}.  }
\end{figure*}

\section{Single origami vertex}

Many of the design strategies for self-folding materials involves a single fold, an array of non-intersecting folds, or an array of folds that intersect only at the boundary of the material \cite{yoon2014functional,py2007capillary,liu2012self,ionov2011soft,stoychev2011self}. From a formal standpoint, we define a fold as a straight line demarcating the boundary between two flat sheets of unbendable, unstretchable material. These sheets, in isolation, are allowed to rotate around the fold, so that the structure behaves mechanically like a simple hinge. If the fold is produced by plastically deforming a piece of material, rather than functioning as a hinge the fold has a preferred angle, and is more precisely called a crease. Herein we shall use the terms interchangeably, since the kinematic motions of a fold and the energetics involved for a crease can be described separately. An important, and arguably defining, characteristic of an origami structure is that it requires that more than one fold meet at a vertex. While each fold individually allows for unrestricted rigid body rotation of a sheet, geometrical constraints arise when several folds coincide at a vertex. These constraints are what provide origami structures with their mechanical novelty, and ultimately are why deployable structures and mechanical metamaterials display exotic and tunable properties.

A vertex of degree $N$ is defined as a point where $N$ straight creases meet. Figure \ref{vertex}A shows the crease pattern for a schematic 6-degree vertex, with sectors defined by planar angles $\alpha_i$. The three-dimensional folded form of this vertex is found by supplying fold angles to each of the creases, subject to the constraints mentioned previously \cite{huffman1976curvature,hull2002modelling}. This procedure is an exercise in spherical trigonometry.

One way to visualize the constraints is to surround each vertex with a sphere and consider the intersection between it and the surface (Fig. \ref{vertex}B). In this construction, the side lengths of the spherical polygon are the angles between adjacent folds, which must remain fixed, and the dihedral fold angles are the internal angles of the polygon on the sphere. Since an $N$-sided polygon has $N-3$ continuous degrees of freedom, each vertex does as well. These $N-3$ degrees of freedom can be thought of, for example, as the angles between a fixed fold and the remaining non-adjacent folds.

Starting with a general vertex containing dihedral angles $f_i$, we use spherical trigonometry to calculate these angles in terms of the $N$-3 degrees of freedom. To calculate $f_1$ we partition the angle into sectors by subdividing the spherical $N$-gon into $N-2$ triangles (Fig. \ref{vertex}C). We label the angles that lead from $f_1$ to $f_i$ as $\ell_i$, where $\ell_1=\alpha_1$ and $\ell_{N-1}=\alpha_N$ are sector angles. All the angles $\alpha_i$ are spherical polygon edges, and since origami structures allow only isometric deformations, these angles are  constant. The $\ell_i$ are the angles subtended by drawing a geodesic on the encapsulating sphere from $f_1$ to $f_i$; expressions for relating the $\ell_i$ to the fold angles $f_i$ are found by using the spherical law of cosines around the vertex \cite{huffman1976curvature}: 

\begin{gather}
f_1=\sum_{i=1}^{N-2}\cos^{-1}\left[\frac{\cos \alpha_{i+1}-\cos\ell_{i+1}\cos\ell_i}{\sin\ell_{i+1}\sin\ell_i}\right],\\
f_2=\cos^{-1}\left[\frac{\cos\ell_{2}-\cos \alpha_{1}\cos\alpha_2}{\sin\alpha_1\sin \alpha_{2}}\right],\\
f_{N}=\cos^{-1}\left[\frac{\cos\ell_{N-2}-\cos \alpha_{N-1}\cos\alpha_N}{\sin\alpha_{N-1}\sin \alpha_{N}}\right],\\
f_i=\cos^{-1}\left[\frac{\cos\ell_{i-2}-\cos \alpha_{i-1}\cos\ell_{i-1}}{\sin\ell_{i-1}\sin \alpha_{i-1}}\right]+\\
\nonumber \cos^{-1}\left[\frac{\cos\ell_{i}-\cos \alpha_{i}\cos\ell_{i-1}}{\sin\ell_{i-1}\sin \alpha_{i}}\right].
\end{gather}

These expressions are essentially all that is required to determine the folding of a single vertex, although the associated solutions are generically multi-valued. These results imply that there are multiple branches of configuration space for any given spherical polygon.

To specify the internal state of each vertex we define an $N-3$ component vector $\mathbf{s}$. Given the internal state of a vertex, all N of the dihedral fold angles are determined, which we collect in the vector $\b{f}(\b{s})$. In practice, computations are vastly simplified by choosing the appropriate degrees of freedom; for example, for a degree 6 vertex of the type displayed in Fig. \ref{vertex}, we choose $\b{s}=\{\ell_3,f_2,f_6\}$, and the fold vector is given by $\b{f}=\{f_1,f_2,f_3,f_4,f_5,f_6\}$.

\section{General lattice theory}

To determine the mechanical properties of an origami tessellation we begin by examining how many vertices are connected together in a crease pattern. When constructing a real piece of origami, artists and designers specify ``mountain" and ``valley" creases in the pattern to encode instructions for how the structure will fold. In our formulation we will treat the crease pattern as a simple connected graph, where each unique crease is an edge that connects two vertices to one another.

\subsection{Kinematically allowed deformations}

\begin{figure}[!]
\includegraphics[width=.5\textwidth]{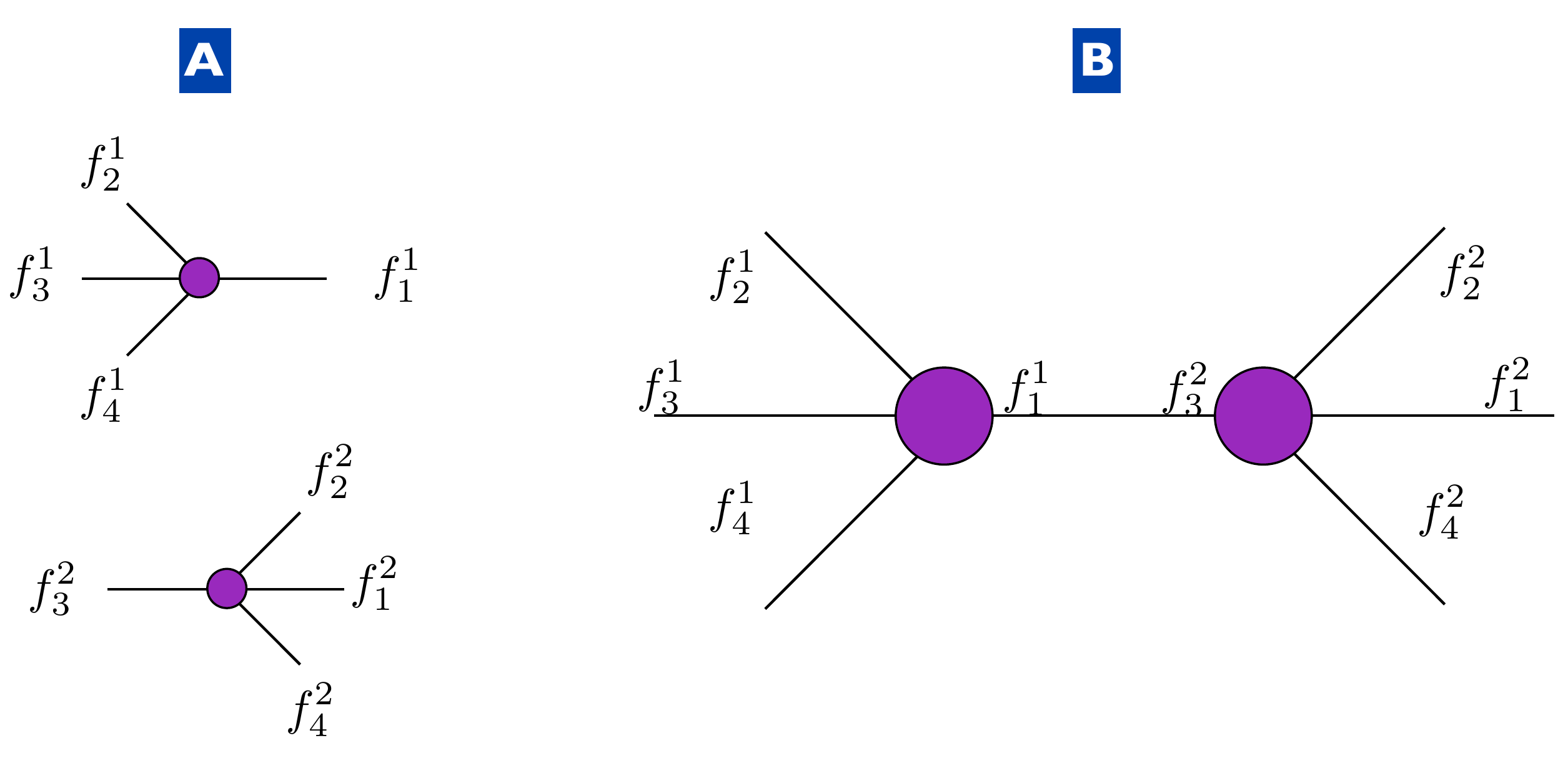}
\caption{\label{constraint_schem}(color online) (A) Two degree four vertices with labeled folds. (B) the graph for the crease pattern consisting of these two vertices contains a single crease that is shared by both vertices. In this case the constraint equation $\b{D}\boldsymbol{\mathcal{F}}=0$ simply becomes the scalar relationship $f_1^1=f_3^2$.}
\end{figure}

In addition to the origami constraints discussed above for a single vertex, joining multiple vertices together generates further constraints on the folds. Consider a crease pattern that consists of $P$ vertices. Each vertex $v_p$, with $p\in\{1,...,P\}$ has $N_p$ folds, collected in the vector $\b{f}^p=(f_1^p f_2^p \cdots f_{N_p}^p)^T$. If we collect all the folds into the vector $\boldsymbol{\mathcal{F}}$, given by

\begin{gather}
\boldsymbol{\mathcal{F}}=\left(\begin{array}{c}f_1^1 \\f_2^1 \\\vdots \\f_{N_1}^1 \\f_1^2 \\\vdots \\f_{N_2}^2 \\\vdots\\f_1^P\\\vdots\\f_{N_P}^P\end{array}\right),
\end{gather}
then we have the following constraint equation for the folds:

\begin{gather}
\label{eq:constraint}
\mathbf{D}\boldsymbol{\mathcal{F}}=0,
\end{gather}
where $\mathbf{D}$ is a sparse rectangular matrix that enforces the condition that if two vertices $v_q,v_p$ are adjacent, and two folds $\mathcal{F}_i,\mathcal{F}_j$ connect $v_q,v_p$, then $\mathcal{F}_i=\mathcal{F}_j$ (see Fig. \ref{constraint_schem} for an example). This constraint enforces the connectivity of the graph, since each unique crease clearly must have a compatible fold angle associated with the vertices that connect it. Each row of $\mathbf{D}$ corresponds to a fold connecting a pair of vertices in the origami tessellation while each column corresponds to a component of $\boldsymbol{\mathcal{F}}$. Analysis of this construction is the essence of origami mechanics, and lies at the heart of the difficulty in determining general properties of tessellations and corrugations. Finding the null vectors of $\boldsymbol{D}$ amounts to finding all of the possible solutions for the fold angles, and thus all of the kinematically allowed motions of the rigid origami. While computational methods have been developed for simulating the kinematics of origami and linkage structures \cite{tachi2010geometric,schenk2011origami,schenk2011folded,schenk2013geometry}, there has been no general analytical study that seeks to identify mechanical properties based solely on the crease and fold patterns.

The functions $\boldsymbol{\mathcal{F}}(\mathbf{s})$ are, in general, nonlinear. To proceed analytically, we expand $\mathbf{s}$ about a state $\mathbf{s}_0$ that solves the constraint equations. That is, if $\boldsymbol{\mathcal{F}}(\mathbf{s}_0)=\boldsymbol{\mathcal{F}}_0$ then $\b{D}\boldsymbol{\mathcal{F}}_0\equiv 0$. A trivial choice for $\b{s}_0$ has every entry identically equal to $\pi$, indicating that the piece of origami is unfolded. The more common, and more interesting, scenario involves a folded state where the values of the internal vector $\b{s}_0$ are known. Assuming that such a state exists, we write $\mathbf{s} = \mathbf{s}_0 + \delta \mathbf{s}$, with $\delta\mathbf{s}$ a small perturbation, and then have 
\begin{equation}
\label{eq:linear}
\mathbf{D} \mathbf{J} \delta \mathbf{s}\equiv\b{R}\delta\b{s}=0,
\end{equation}
where the Jacobian of the fold angles for each vertex $\mathbf{J} \equiv \partial\boldsymbol{\mathcal{F}}/\partial \mathbf{s}|_{\mathbf{s}_0}$ is a block diagonal matrix defining the small deviations from the ``ground state" $\b{s}_0$, and $\b{R}$ is a rigidity matrix that informs on the infinitesimal isometric deformations of the origami structure \cite{hutchinson2006structural,kane2014topological}. This formulation is convenient since it separates the effects of the crease pattern topology (contained entirely in $\b{D}$) from the constrained motion of a single vertex (contained entirely in $\b{J}$). We can thus solve for each of these matrices individually. 

To find $\mathbf{D}$, we first exploit the periodicity of the lattice to decompose the vector $\boldsymbol{\mathcal{F}}$ and matrix $\b{D}$ in a Fourier basis, such that $\boldsymbol{\mathcal{F}}=\sum_{n,m}e^{i \b{q}\cdot\b{x}}\boldsymbol{\mathcal{F}}_q+c.c.$. Here $\b{q}$ is a two-dimensional wave-vector and $\b{x}=n \b{a}_1+m\b{a}_2$ is the 2D position vector of the fundamental unit cell on the crease pattern lattice, where $(n,m)$ indexes this position in terms of the lattice vectors $\b{a}_{1,2}$. Since $\boldsymbol{\mathcal{F}}\approx\b{J}\delta\b{s}$ and $\b{J}$ is independent of the lattice position, we also have $\delta\b{s}=\sum_{n,m}e^{i \b{q}\cdot\b{x}}\delta\b{s}_q+c.c.$, where $\boldsymbol{\mathcal{F}}_q=\b{J}\delta\b{s}_q$. In this representation the constraints given in Eq. \ref{eq:constraint} are 

\begin{gather}
\b{D}(\b{q})\boldsymbol{\mathcal{F}}_{\b{q}}=\b{D}(\b{q})\b{J}\delta\b{s}_q=0
\end{gather}

 Now, instead of a matrix operation over all the vertices, the size of $\b{D}(\b{q})$ is vastly simplified. For a pattern with $p$ distinct vertices per unit cell, each of degree $N_p$, $\b{D}(\b{q})$ is a $\left(\sum_{i=1}^p{(N_i/2)}\times\sum_{i=1}^p{N_i}\right)$ matrix. In Fourier space, $\b{D}(\b{q})$ is the complex-valued constraint matrix for the graph of the unit cell vertices and folds. Specifically, each fold of the unit cell is represented by a row in $\mathbf{D}(\mathbf{q})$ having only two nonzero entries. Those entries all have the form $\pm e^{i \mathbf{q} \cdot \mathbf{a}_1},\pm e^{i \mathbf{q} \cdot \mathbf{a}_2},\pm 1$, depending on whether the fold connects to an adjacent unit cell along $\mathbf{a}_{1,2}$ or is internal to the unit cell.

The formulation in terms of the matrix $\mathbf{R}(\mathbf{q})$ is completely general for any origami tessellation. The rectangular matrix $\mathbf{D}(\mathbf{q})$ carries all of the topological information regarding the fold network, while the Jacobian $\b{J}$ carries the information about the type of vertex that has been specified. $\mathbf{J}$ will be block diagonal with one block for each vertex of a unit cell, but does not depend on $\mathbf{q}$ for a regular tessellation.

\subsection{Origami energetics}

While the $\mathbf{R}$ matrix determines the kinematically isometric deformation to leading order, these constraints are generally not the end of the story for real materials. Creases in folded paper, thermoresponsive gels with programmed folding angles, and elastocapillary hinges all balance energetic considerations with geometric constraints. In many cases these creases and hinges act as torsional springs, while the bending of faces have additional elastic energy content \cite{silverberg2014using,lechenault2014mechanical,silverberg2015origami}. 

The energy associated with the entire structure may be written, to quadratic order in the dihedral vectors, as

\begin{gather}
\mathcal{E}=\frac{1}{2}\left(\boldsymbol{\mathcal{F}}-\boldsymbol{\mathcal{F}}_0\right)^T\mathcal{A}\left(\boldsymbol{\mathcal{F}}-\boldsymbol{\mathcal{F}}_0\right),
\end{gather}
where $\mathcal{A}$ is a general stiffness matrix and $\boldsymbol{\mathcal{F}}_0$ is a reference fold angle. For linear response this is the most generic form for the energy. In the simplest of cases $\mathcal{A}$ is constant over the lattice and diagonal with respect to $\boldsymbol{\mathcal{F}}$; this models each crease as a torsional spring with uniform spring constant \cite{wei2013geometric,silverberg2014using,silverberg2015origami}. Small amplitude response is found by examining the origami structure near the ground state, that is, when $\boldsymbol{\mathcal{F}}=\boldsymbol{\mathcal{F}}_0$. When the energy is expanded about the ground state $\mathcal{E}_0$ we find 

\begin{gather}
\mathcal{E}=\mathcal{E}_0+\frac{1}{2}\delta\b{s}^T \b{J}^T\mathcal{A}\b{J}\delta\b{s},
\end{gather}
or in the Fourier decomposition,

\begin{gather}
\label{energy}
\mathcal{E}=\frac{LW}{2}\sum_{\mathbf{q}} \delta \mathbf{s}^\dagger_\b{q} \mathcal{M}\delta \mathbf{s}_\b{q},
\end{gather}
where $L$ is the length of the tessellation in the $\b{a}_1$ direction, $W$ is the width in the $\b{a}_2$ direction, and $\mathcal{M}=\b{J}^T\mathcal{A}\b{J}$ is a matrix operator that is independent of wavenumber. Since the nullspace of $\b{R}(\b{q})$ will determine the modes of deformation, the solution to this problem lies in finding the kinematically allowed deformations, and then any energetic description will simply involve a change of basis to a system of deformations that diagonalize the operator $\mathcal{M}$.

\section{Miura-ori}

\begin{figure}[!]
\includegraphics[width=.5\textwidth]{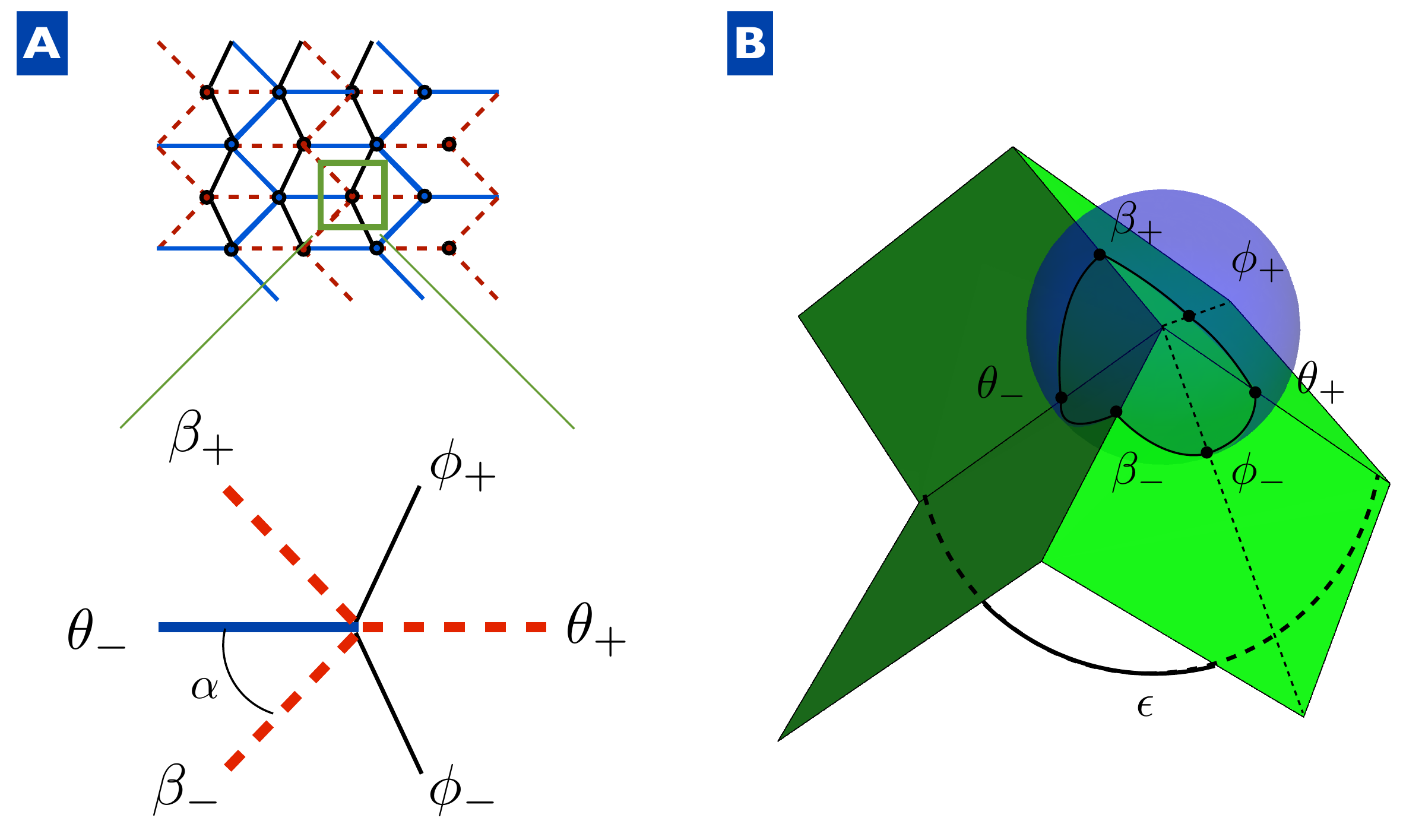}
\caption{\label{lattice_schem} (color online) (A) While the crease pattern of a Miura-ori generally introduces only four folds per vertex, the bending of faces acts to allow two extra folds per vertex, so the crease pattern we consider is a triangulated lattice. At each vertex the dihedral angles contained in $\b{f}$ are determined by specifying the state vector $\b{s}$ and satisfying the geometric constraints. (B) Single vertex origami with enclosing sphere to visualize the constraints between $\b{f}$ and $\b{s}$.}
\end{figure}

\begin{figure*}[!]
\includegraphics[width=.95\textwidth]{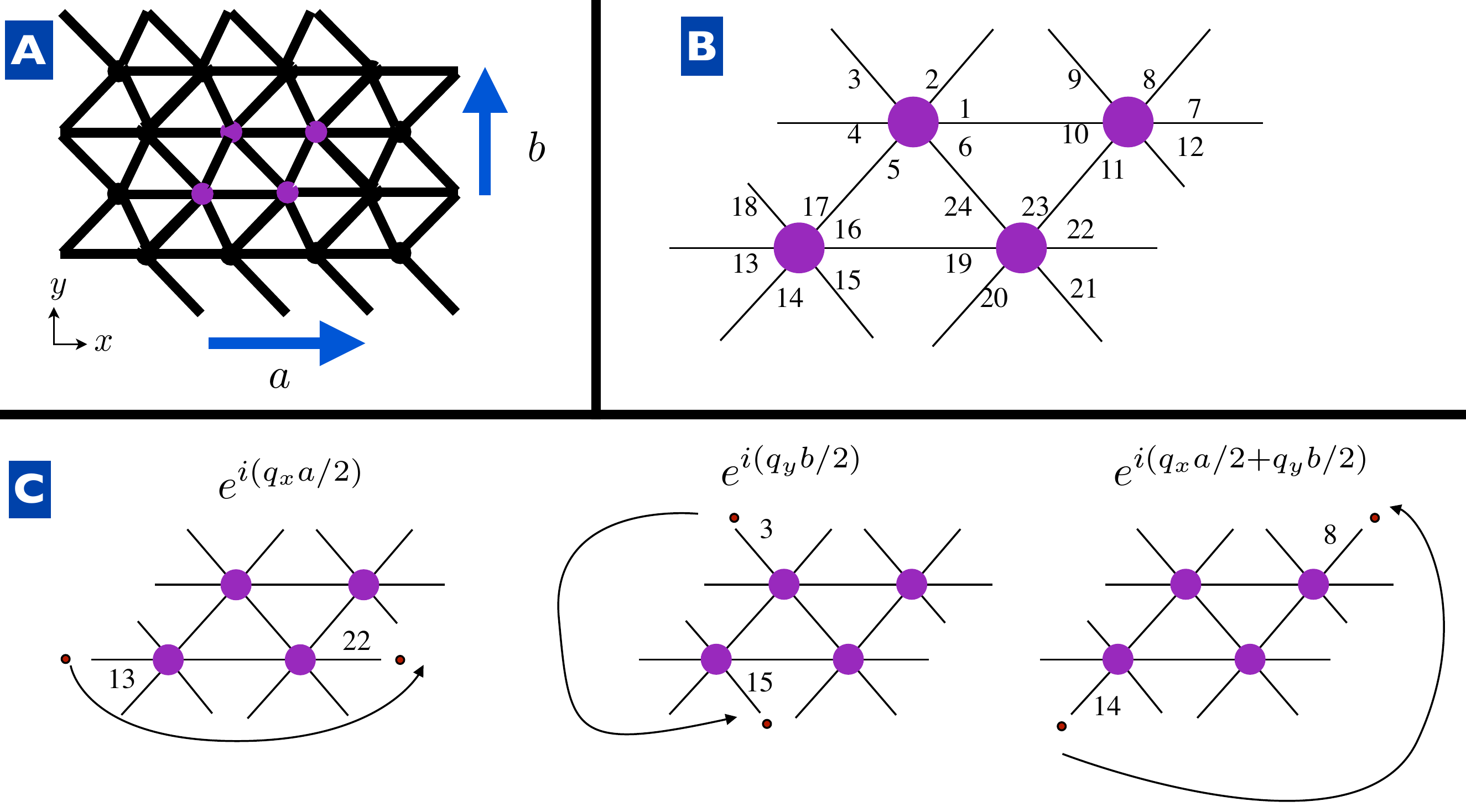}
\caption{\label{dmatrix} (color online) (A) Miura-ori, without the assignment of mountain/valley folds, has a simple directed graph structure with a unit cell composed of four vertices. By tessellating these four vertices the entire pattern emerges. Note that the tessellation is rectangular, with lattice vectors $\b{a}_1=a\b{\hat{x}}$ and $\b{a}_2=b\b{\hat{y}}$. (B) Each vertex has six folds, labelled in the fashion shown here. (C) In Fourier space, translations associated with connecting these folds together throughout the tessellation merely amounts to a phase factor associated with the appropriate wave number and lattice vector. Left: Translating in the x direction. Middle: Translating in the y direction. Right: Connecting the extra folds involves a diagonal translation across the unit cell. Note that the five internal folds have a phase factor identically equal to one.}
\end{figure*}

As an example of this formulation, we consider inhomogeneous deformations of a particular origami metamaterial, the Miura-ori. First introduced as a framework for a deployable surface, the design appears often in nature, from plant leaves \cite{mahadevan2005self} to animal viscera \cite{shyer2013villification}. Additionally, theoretical calculations and experiments have suggested the Miura-ori as a canonical, origami-based, auxetic metamaterial \cite{schenk2011folded,schenk2011origami,schenk2013geometry,wei2013geometric,silverberg2014using}. Its ubiquity may be related to its simplicity: the Miura-ori is determined from a single crease angle $\alpha$ and the mountain/valley assignments of the pattern shown in Fig. \ref{lattice_schem}. Conventional origami mathematics considers that each Miura-ori vertex is degree four, and thus there is only one degree of freedom. However, casual experimentation with a real Miura-ori quickly demonstrates that it has far more than one degree of freedom, indicating an array of ``soft modes" enabled by the bending of the individual faces. This breakdown of the assumptions of mathematical origami is well known, and there are many crease patterns that are mathematically impossible to fold that can in fact be done with little effort \cite{demaine2011non}. To incorporate these extra degrees of freedom into Miura-ori, we assume that there are two extra folds per vertex to account for face bending. While in the extreme case of the creases being perfectly rigid these extra folds would actually take the form of stretching ridges \cite{witten2007stress}, many real applications involve fabrication processes that will allow the face to be well approximated as perfect bending. Each unit cell in the tessellation has four six-valent vertices (Fig. \ref{lattice_schem}) so there are $12$ degrees of freedom per unit cell. In this example the fold vector for the $i^{th}$ vertex is given by $\b{f}^i=(\theta^i_+,\phi^i_+,\beta^i_+,\theta^i_-,\beta^i_-,\phi^i_-)^T$, and the vector $\boldsymbol{\mathcal{F}}=(\b{f}^1\,\, \b{f}^2 \,\,\b{f}^3\,\, \b{f}^4)^T$. There are three degrees of freedom per vertex that define the internal state $\b{s}$, which we parameterize using three angles: $\epsilon$, the angle between folds labeled $\theta_\pm$ in Fig. \ref{lattice_schem}, and the angles $\phi_\pm$ representing the bending of the faces. Using the geometric relationships between the angles \cite{huffman1976curvature}, we find the general nonlinear relationship for a single vertex, and then expand  about the ground state $\b{s}_0=\{\epsilon + \delta \epsilon,\pi + \delta \phi_+,\pi+\delta \phi_-\}$ to find the matrix $\mathbf{J}$; here $\epsilon\in[\pi-2\alpha,\pi+2\alpha]$. This expansion naturally follows from assuming that the faces are nearly flat and that the Miura-ori has been folded into the standard configuration. The Jacobian $\b{J}=diag\left(\begin{array}{cccc}\b{J}_0 & -\b{J}_0 & \b{J}_0 & -\b{J}_0\end{array}\right)$ is a $24 \times 12$ diagonal block matrix formed from four identical blocks,
\begin{equation}
\mathbf{J}_0 = \left(
\begin{array}{ccc}
A & C & C\\
0 & 1 & 0\\
B & C & 0\\
-A & 0 & 0\\
B & 0 & C\\
0 & 0 & 1
\end{array}
\right)
\end{equation}
where 

\begin{gather}
A = \cos \alpha \csc(\epsilon/2)/\sqrt{\sin^2 (\epsilon/2) - \cos^2 \alpha},\\
B = \sin(\epsilon/2)/ \sqrt{\sin^2 (\epsilon/2) - \cos^2 \alpha},\\
C = \csc (\alpha/2)/2.
\end{gather}

To calculate the constraint matrix, we note that there are 12 unique folds per unit cell so that $\mathbf{D}(\mathbf{q})$ is a $12 \times 24$ rectangular matrix. It has a row for each bond in Fig. \ref{dmatrix} with two nonzero columns indicating which folds of each vertex are interconnected. For internal folds the constraint matrix has a value of $\pm 1$, while folds that leave the unit cell have a phase factor associated with it. The full matrix is given by:

\begin{widetext}
\begin{gather}
\b{D}^T(\b{q})=\left(
\begin{array}{cccccccccccc}
 1 & 0 & 0 & 0 & 0 & 0 & 0 & 0 & 0 & 0 & 0 & 0 \\
 0 & e^{\frac{i q_y}{2}} & 0 & 0 & 0 & 0 & 0 & 0 & 0 & 0 & 0 & 0 \\
 0 & 0 & e^{\frac{i q_y}{2}} & 0 & 0 & 0 & 0 & 0 & 0 & 0 & 0 & 0 \\
 0 & 0 & 0 & -e^{-\frac{i q_x}{2}} & 0 & 0 & 0 & 0 & 0 & 0 & 0 & 0 \\
 0 & 0 & 0 & 0 & -1 & 0 & 0 & 0 & 0 & 0 & 0 & 0 \\
 0 & 0 & 0 & 0 & 0 & -1 & 0 & 0 & 0 & 0 & 0 & 0 \\
 0 & 0 & 0 & e^{\frac{i q_x}{2}} & 0 & 0 & 0 & 0 & 0 & 0 & 0 & 0 \\
 0 & 0 & 0 & 0 & 0 & 0 & e^{\frac{i q_x}{2}+\frac{i q_y}{2}} & 0 & 0 & 0 & 0 & 0 \\
 0 & 0 & 0 & 0 & 0 & 0 & 0 & e^{\frac{i q_y}{2}} & 0 & 0 & 0 & 0 \\
 -1 & 0 & 0 & 0 & 0 & 0 & 0 & 0 & 0 & 0 & 0 & 0 \\
 0 & 0 & 0 & 0 & 0 & 0 & 0 & 0 & -1 & 0 & 0 & 0 \\
 0 & 0 & 0 & 0 & 0 & 0 & 0 & 0 & 0 & e^{\frac{i q}{2}} & 0 & 0 \\
 0 & 0 & 0 & 0 & 0 & 0 & 0 & 0 & 0 & 0 & -e^{-\frac{i q}{2}} & 0 \\
 0 & 0 & 0 & 0 & 0 & 0 & -e^{-\frac{i q_x}{2}-\frac{i q_y}{2}} & 0 & 0 & 0 & 0 & 0 \\
 0 & 0 & -e^{-\frac{i q_y}{2}} & 0 & 0 & 0 & 0 & 0 & 0 & 0 & 0 & 0 \\
 0 & 0 & 0 & 0 & 0 & 0 & 0 & 0 & 0 & 0 & 0 & 1 \\
 0 & 0 & 0 & 0 & 1 & 0 & 0 & 0 & 0 & 0 & 0 & 0 \\
 0 & 0 & 0 & 0 & 0 & 0 & 0 & 0 & 0 & -e^{-\frac{i q_x}{2}} & 0 & 0 \\
 0 & 0 & 0 & 0 & 0 & 0 & 0 & 0 & 0 & 0 & 0 & -1 \\
 0 & -e^{-\frac{i q_y}{2}} & 0 & 0 & 0 & 0 & 0 & 0 & 0 & 0 & 0 & 0 \\
 0 & 0 & 0 & 0 & 0 & 0 & 0 & -e^{-\frac{i q_y}{2}} & 0 & 0 & 0 & 0 \\
 0 & 0 & 0 & 0 & 0 & 0 & 0 & 0 & 0 & 0 & e^{\frac{i q_x}{2}} & 0 \\
 0 & 0 & 0 & 0 & 0 & 0 & 0 & 0 & 1 & 0 & 0 & 0 \\
 0 & 0 & 0 & 0 & 0 & 1 & 0 & 0 & 0 & 0 & 0 & 0 \\
\end{array}
\right)
\end{gather}
\end{widetext}

\subsection{Bulk deformation}

\begin{figure*}[!]
\includegraphics[width=.75\textwidth]{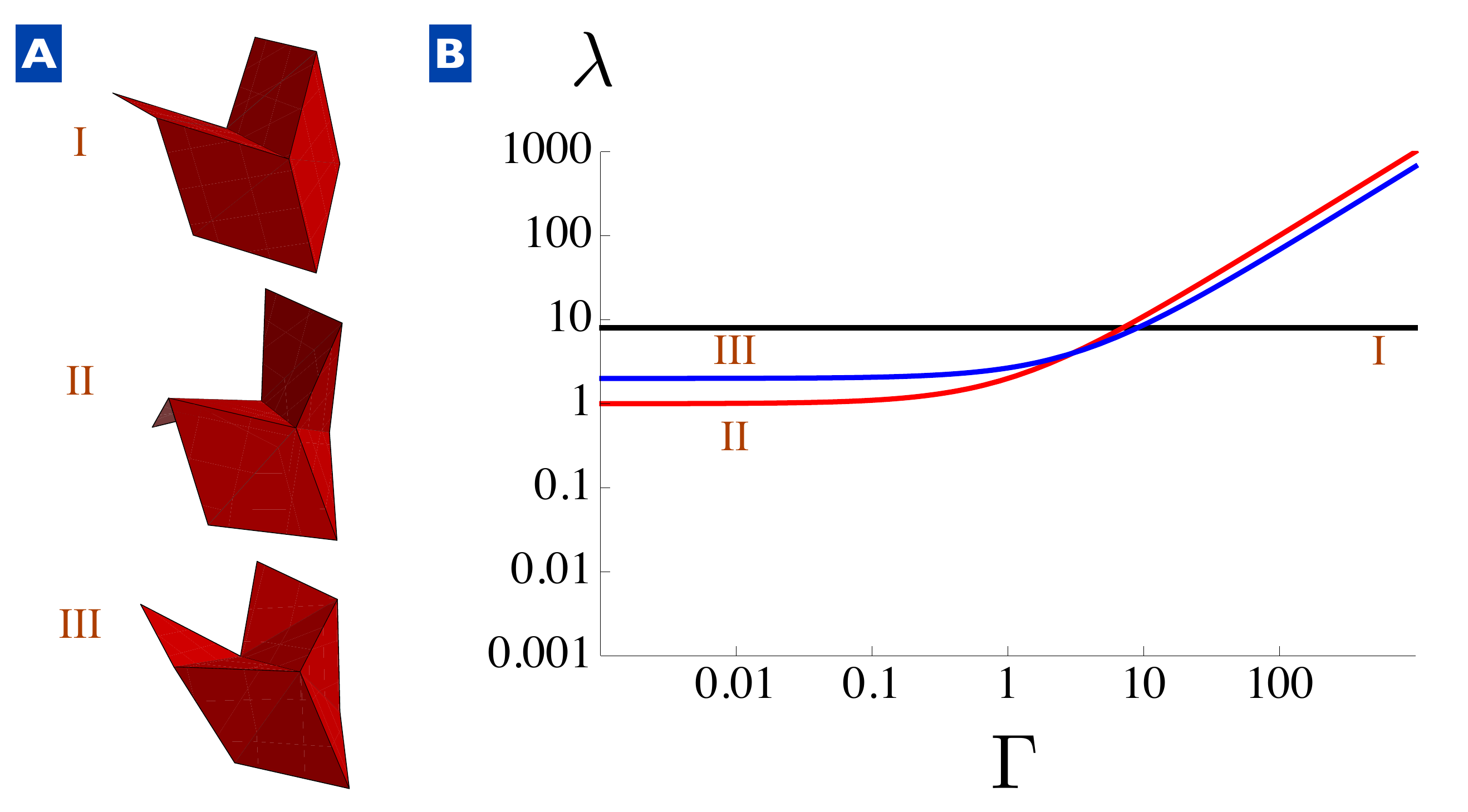}
\caption{\label{embedding} (color online) Shapes and energy eigenvalues for the three uniform modes for $\epsilon=\pi/2$ and $\alpha=\pi/3$. (A) The three uniform null vectors correspond to a uniform mode (I), a twisting mode (II), and a saddle mode (III). These are identical to the modes determined numerically in previous studies \cite{schenk2011origami,schenk2013geometry}. (B) Eigenvalues associated with each of the three bulk modes as a function of face stiffness $\Gamma$. Note that over a wide range the softest mode is the twisting mode (II), since it involves purely face bending.}
\end{figure*}

The combination $\mathbf{D}(\mathbf{q}) \mathbf{J}$ is square such that Eq. (\ref{eq:linear}) has a nontrivial solution whenever $\textrm{det} \left[ \mathbf{D}(\mathbf{q}) \mathbf{J}\right] = 0$. We non-dimensionalize the wavenumber by the physical lengths of the lattice vectors such that $q_x\rightarrow q_x a$ and $q_y\rightarrow q_y b$, and the resulting dispersion relation is
\begin{equation}\label{eq:disp1}
\frac{\cos^2 \alpha}{\sin^4(\epsilon_0/2)} \sin^2(q_x/2) + \sin^2(q_y/2) = 0.
\end{equation}

The only real solution to this equation is $\b{q}=0$, indicating that an infinite origami tessellation does not admit spatially inhomogeneous solutions; only uniform deformations are allowed. The nullspace of $\b{R}$ is three dimensional here, corresponding to three uniform deformation modes of the Miura-ori. These zero modes are given by the vectors $\b{\Psi}_i$:

\begin{gather}
\b{\Psi}_I=\left(\begin{array}{c}1 \\0 \\0 \\1 \\0 \\0 \\1 \\0 \\0 \\1 \\0 \\0\end{array}\right),\,\,\,\,\, \b{\Psi}_{II}=\left(\begin{array}{c}0 \\-1 \\1 \\0 \\-1 \\1 \\0 \\-1 \\1 \\0 \\-1 \\1\end{array}\right),\,\,\,\,\, \b{\Psi}_{III}=\left(\begin{array}{c}-2\frac{C}{A} \\1 \\1 \\0 \\1 \\1 \\-2\frac{C}{A} \\1 \\1 \\0 \\1 \\1\end{array}\right)
\end{gather}

These infinitesimal deformations of the unit cell correspond to a uniform contraction, a twisting mode, and a saddle-like deformation, respectively (see Fig. \ref{embedding}).

To describe the kinematics of deformation, all we require are the null vectors of the constraint equations, but for examining energy associated with the creases we need to calculate the eigenvalues of the matrix $\mathcal{M}=\b{J}^T\mathcal{A}\b{J}$. In general, it is not unreasonable to assume that a creased and folded Miura-ori will have a crease stiffness $k$ that is approximately equal for all patterned creases, but the energy scale for bending of the faces will depend on the material properties of the structure \cite{silverberg2014using}. The energy for bending can be treated as an effective torsional spring constant $k_b$, and thus the energy can be written in terms of the ratio $k_b/k\equiv\Gamma$. Non-dimensionalizing the energy by $kL_xL_y$, we find the energy eigenvalues $\lambda$ in terms of the null vectors. Decomposing the internal variable deformation $\delta \b{s}=\sum_i a_i\b{\psi}_i$, where  $\b{\psi}_i=\b{\Psi}_i/|\b{\Psi}_i|$ is the normalized null vector with $i\in \{I,II,III\}$, we write Eq. \ref{energy} as

\begin{gather}
E=\frac{L_xL_y}{2}\b{a}^T\b{M}\b{a},\\
\mathbf{a} = \left(
\begin{array}{c}
a_I\\
a_{II}\\
a_{III}
\end{array}\right),\\
\mathbf{M} = \left(
\begin{array}{ccc}
\psi^T_I\mathcal{M}\psi_I & \psi^T_I\mathcal{M}\psi_{II} & \psi^T_I\mathcal{M}\psi_{III}\\
\psi^T_{II}\mathcal{M}\psi_I & \psi^T_{II}\mathcal{M}\psi_{II} & \psi^T_{II}\mathcal{M}\psi_{III}\\
\psi^T_{III}\mathcal{M}\psi_I & \psi^T_{III}\mathcal{M}\psi_{II} & \psi^T_{III}\mathcal{M}\psi_{III}
\end{array}\right).
\end{gather}

Each matrix element of $\b{M}$ represents overlaps between the null vectors $\b{\psi}_i$ and the energy matrix $\mathcal{M}$; only in exceptional circumstances will $\b{M}$ be diagonal in the null basis. In general it is given by

\begin{gather}
\b{M}=\left(
\begin{array}{ccc}
 2 \left(A^2+B^2\right) & 0 & \frac{\sqrt{2} (A-B) B C}{
   \sqrt{C^2+A^2}} \\
 0 & C^2+\Gamma  & 0 \\
\frac{\sqrt{2} (A-B) B C}{
   \sqrt{C^2+A^2}}& 0 &
   \frac{\Gamma  A^2+\left(3 A^2-2 B A+2 B^2\right)
   C^2}{A^2+C^2} \\
\end{array}
\right)
\end{gather}

An example for when $\b{M}$ is diagonal is given by $\alpha=\pi/3,\epsilon=\pi/2$ (see Fig. \ref{embedding}), for which $\b{M}$ becomes:

\begin{gather}
\mathbf{M}=\left(\begin{array}{ccc}
8 & 0 & 0\\
0 & 1+\Gamma & 0\\
0 & 0 & \frac{2}{3}(3+\Gamma)
\end{array}\right)
\end{gather}

Note that the for this particular combination of parameters the uniform expansion mode has a flat stiffness over all ranges of $\Gamma$ since there is no face bending for this deformation. In the regime where face bending is relatively inexpensive ($\Gamma\ll1$), the out of plane deformation modes are correspondingly softer than the uniform deformation. These results are in agreement with previous numerical research done on the structural mechanics of Miura-ori \cite{schenk2011origami,schenk2011folded,schenk2013geometry}. Should other values of $(\alpha,\epsilon)$ be chosen, the energy matrix is not necessarily diagonal, and thus eigensolutions mix the null vectors.

\begin{figure*}[!]
\includegraphics[width=.75\textwidth]{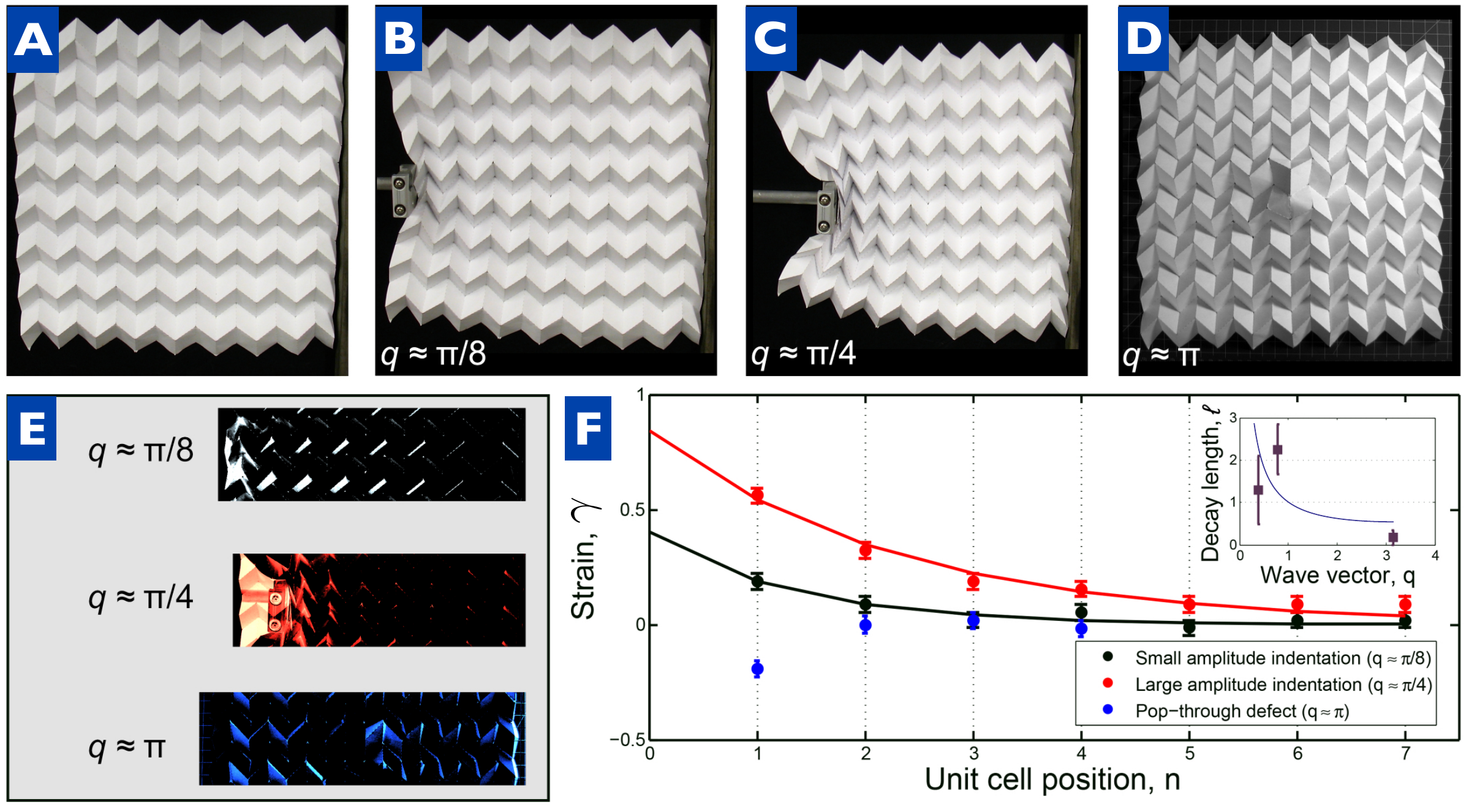}
\caption{\label{experiment} (color online) Experimental observations of deformation localization in an $8\times8$
Miura-ori tessellation.  (A) An undeformed Miura-ori shows a regular
periodic pattern.  Under (B) small deformations, (C) large
deformations, and (D) in the presence of a ``pop-through defect" (PTD) \cite{silverberg2014using},
the lattice distorts to accommodate the induced strain.  (E)
Qualitatively, the amount of deformation localization can be easily
seen by a simple image subtraction between the deformed and undeformed
state.  (F) Measuring strain along the horizontal axis as a function of unit cell
position $n$ relative to the location of the disturbance
shows a rapid decay for all three scenarios (points). For small and
large amplitudes, the decays can be readily fit to an
exponential function with decay length $\ell$ (red/upper and black/lower lines),
whereas for a PTD, the decay length can be estimated
to within  $100\%$.  Because the PTD induces an
extensional distortion rather than a compression, the strain is
oppositely signed.  (Inset) Plotting the decay length against an
approximate measure of the distortion wave vector $q$ shows the larger
wave vector decays much more rapidly than the shorter wave vectors.
Within errorbars, this measurement is
consistent with an inverse relationship between decay length and wave
vector. The solid line is the theoretical prediction from Eq. \ref{length} for $\epsilon=\pi/2$ and $\alpha=\pi/3$.}
\end{figure*}

\subsection{Inhomogeneous deformation}

For a finite tessellation, the deformation is fundamentally different, since some folds reach the boundary and, consequently, do not yield constraints. Since the tessellation mechanics are determined by the allowable deformations, which are determined by the constraint equations, the presence of free boundaries allows much more flexibility, and the Miura-ori develops additional degrees of freedom. These localized ``edge states" are reminiscent of evanescent waves in electromagnetism, boundary layers in elastic lattices \cite{phani2008elastic}, and Rayleigh surface waves \cite{strutt1885waves}. Letting $q_x\equiv q$ (where $q$ is real),
Eq. (\ref{eq:disp1}) yields $q_y = \pm i \kappa(q)$, where deformations decay away from the boundaries of constant $y$ with a length scale $\ell\equiv1/\kappa(q)$, with

\begin{gather}
\label{length}
\ell(q)=\frac{1}{2 |\sinh^{-1} [\frac{\cos\alpha\sin(q/2)}{\sin^2\epsilon/2}]|}.
\end{gather}

This localization length is readily observed in deformation experiments on Miura-ori sheets (see Fig. \ref{experiment}). Using laser-cut sheets of paper, an $8\times8$ Miura-ori is constructed by folding the whole sheet using a planar angle of $\alpha=\pi/3$ into the ground state given by $\epsilon=\pi/2$ (Fig. \ref{experiment}A). Inhomogeneous deformations are created using both an external indenter to apply a displacement  (Fig. \ref{experiment}B,C) and by placing reversible ``pop-through defects" (Fig. \ref{experiment}D) \cite{silverberg2014using}.  The strain $\gamma_n$ at each unit cell $n$ is measured such that $\gamma_n = \Delta w_n /\bar{w}$, where $\Delta w_w$ is the change in width of the $n^{th}$ cell and $\bar{w}$ is the average width for an undisturbed cell. As shown in Fig.\ref{experiment}, the strain decays exponentially away from the indenter with a decay length that is consistent (within error) with our theoretical predictions.

To examine these deformation modes more quantitatively, we return to the ``dispersion relation" given by Eq. \ref{eq:disp1}. There are two possible solutions to Eq. \ref{eq:disp1}, corresponding to different decay directions, and thus the null space of $\mathbf{R}$ corresponding to each of these branches is two dimensional. We decompose $\delta \b{s}(\b{x})$ into a sum of upward (in $y$) decaying and downward decaying modes,

\begin{gather}
\delta \b{s}=e^{i q x}\Big(\left[u_1\boldsymbol{\chi}_1e^{-k(q) y}+u_2\boldsymbol{\chi}_2e^{-k(q) y}\right]+\\
\nonumber \left[d_1\boldsymbol{\eta}_1e^{k(q) y}+d_2\boldsymbol{\eta}_2e^{k(q) y}\right]\Big)+c.c.
\end{gather}

The vectors $\boldsymbol{\chi}_{1,2}$ correspond to the upward decaying modes, while $\boldsymbol{\eta}_{1,2}$ the downward decaying modes. Note that, since the values of the angles must be real, $\boldsymbol{\eta}_{1,2}(q)=\boldsymbol{\bar{\chi}}_{1,2}(-q)$. In the long-wavelength limit, i.e. $q\ll1$, we have:

\begin{gather}
\boldsymbol{\chi}_1=\left(\begin{array}{c}-\frac{CA q+CB(q-2i)}{A B}\\0\\2q\\\frac{-CA q+CB
  (q+2 i)}{A B}\\0\\0\\\frac{CAq-CB (q-2 i)}{A B}\\0\\0\\\frac{CAq+CB (q+2 i)}{A B}\\0\\2q\end{array}\right), \,\,\,\,\,\, \boldsymbol{\chi}_2=\left(\begin{array}{c}-\frac{2 C}{A}\\0\\0\\-\frac{2 C}{A}\\-2 i q\\0\\-\frac{2 C}{A}\\-2 i
   q\\0\\-\frac{2 C}{A}\\0\\0\end{array}\right)
\end{gather}

The nullspace, and thus the number of elementary excitations, for a finite-sized Miura-ori is actually different than for the limit $q\rightarrow0$. While this may seem counter-intuitive, the nature of the null vectors is inherently chiral, as indicated by the decomposition into upward and downward decaying solutions. At $q=0$, the dimensionality of the nullspace is smaller because there is no distinction between handedness for uniform deformation.

\begin{figure*}[!]
\includegraphics[width=.85\textwidth]{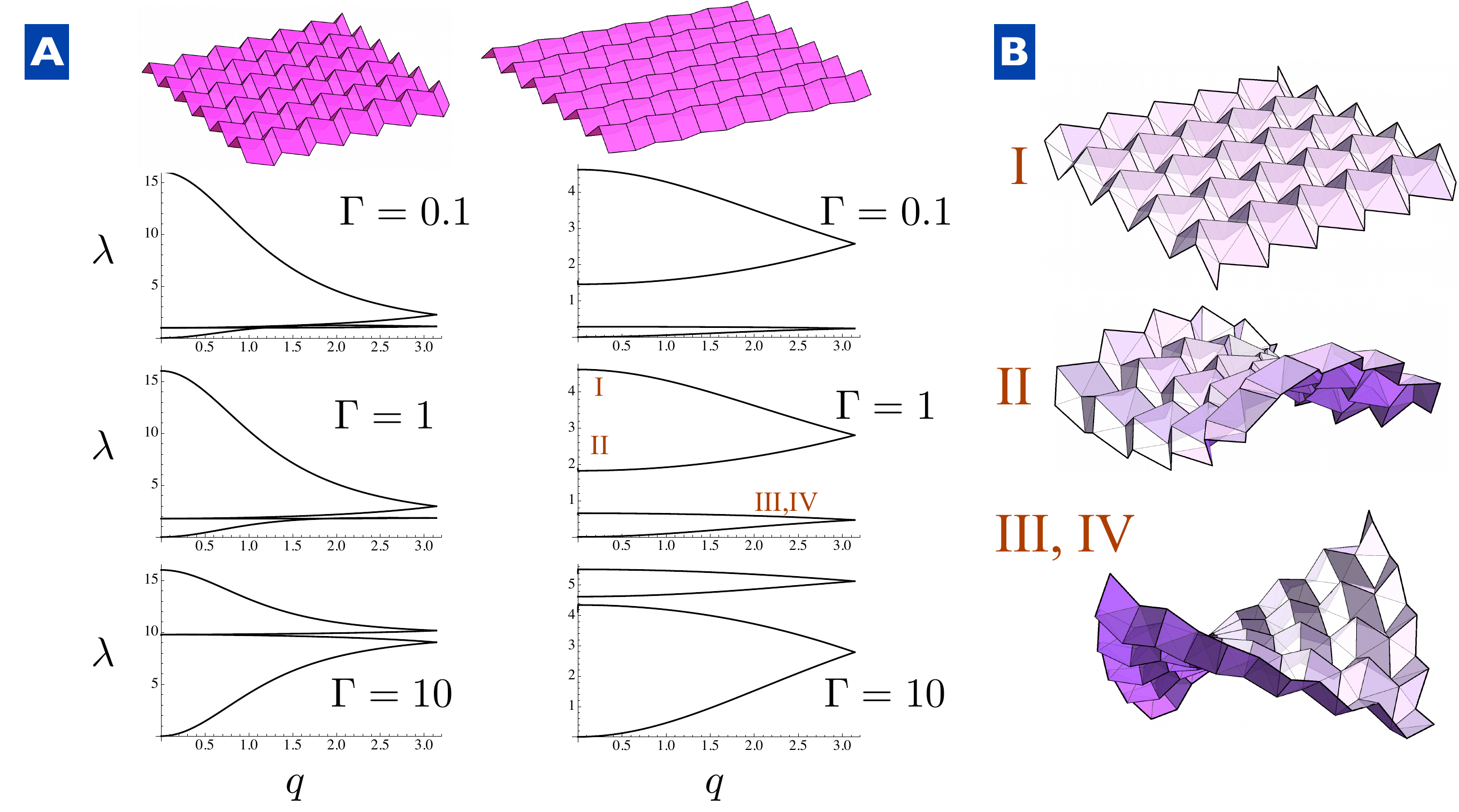}
\caption{\label{results} (color online) Eigenvalues and mode shapes as a function of wavenumber for a given $\Gamma$. (A) Left: Mode structure for i) $\Gamma=0.1$, ii) $\Gamma=1$, and iii)$\Gamma=10$, with $\epsilon=\pi/2$ and $\alpha=\pi/3$. At long wavelengths the saddle mode $\mathcal{I}$ is the stiffest for a wide range of $q$, since it involves both bending of the faces and deformation of the angles away from the reference state. Right: Mode structure for i) $\Gamma=0.1$, ii) $\Gamma=1$, and iii) $\Gamma=10$, with $\epsilon=\pi/2$ and $\alpha=9\pi/20$. (B) Visualization of the basic modes for $q=\pi/6$.}
\end{figure*}

\subsection{Miura-ori's ``soft modes"}

The vectors $\boldsymbol{\chi}_{1,2}$ govern the kinematic deformations of Miura-ori, giving the possible solutions to the constraint equations. For a tessellation with an associated torsional spring energy at each crease, the energy density per mode may be written in Fourier space as
\begin{equation}
\mathcal{E} = \frac{L_x L_y}{2} \mathbf{c}^\dagger(q) \mathbf{H}(q) \mathbf{c}(q),
\end{equation}
where
\begin{equation}
\mathbf{c}(q) = \left(
\begin{array}{c}
u_1(q)\\
u_2(q)\\
d_1(q)\\
d_2(q)
\end{array}\right),
\end{equation}
and $\mathbf{H}$ is the $2\times2$ Hermitian block matrix ,
\begin{equation}
\mathbf{H} = \left(
\begin{array}{cc}
\mathbf{H}_0 & \mathbf{H}_1\\
\mathbf{H}_1^\dagger & \mathbf{H}_0^\dagger
\end{array}\right).
\end{equation}

The two independent blocks of $\mathbf{H}$ are given by
\begin{equation}
\mathbf{H}_0 = \left(
\begin{array}{cc}
{\boldsymbol{\chi}}_1^{\dagger} \mathcal{M} \boldsymbol{\chi}_1 & {\boldsymbol{\chi}}_1^{\dagger} \mathcal{M} \boldsymbol{\chi}_2\\
{\boldsymbol{\chi}}_2^{\dagger} \mathcal{M} \boldsymbol{\chi}_1 & {\boldsymbol{\chi}}_2^{\dagger} \mathcal{M} \boldsymbol{\chi}_2\end{array}
\right).
\end{equation}
and
\begin{equation}
\mathbf{H}_1 = \left(
\begin{array}{cc}
{\boldsymbol{\chi}}_1^{\dagger} \mathcal{M} \boldsymbol{\eta}_1 & {\boldsymbol{\chi}}_1^{\dagger} \mathcal{M} \boldsymbol{\eta}_2\\
{\boldsymbol{\chi}}_2^{\dagger} \mathcal{M} \boldsymbol{\eta}_1 & {\boldsymbol{\chi}}_2^{\dagger} \mathcal{M} \boldsymbol{\eta}_2
\end{array}
\right).
\end{equation}

For finite wavenumber there are four modes of deformation. Typical eigenvalues of $\mathbf{H}(q)$ are shown in Fig. \ref{results}. The largest two eigenvalues are typically associated with changing $\epsilon$, since there is an energetic cost even for very small $\Gamma$. The typically smallest two eigenvalues correspond to twisting mode and a fourth mode that has no analogue in the zero wavenumber case. This mode has a qualitative shape that is similar to the twisting mode, and an energy that vanishes as $q\rightarrow 0$, much like an acoustic mode in a crystal. Previous analyses of inhomogeneous deformations have not found this mode, which we identify here as arising from the breaking of continuous symmetry when a boundary is added to one side of the tessellation. The acoustic mode corresponds to an antisymmetric combination of upward and downward decaying modes; consequently, as $q$ becomes smaller, the change in fold angles associated with the combination cancel, and only three modes appear at $q=0$. 

The modes that are softest depend not only on the stiffness of face bending, but on the ground state defined by $\epsilon_0$ (see Fig. \ref{results}). This stiffness dependence is in accord with the previously predicted anisotropic in-plane stiffness response \cite{schenk2013geometry,wei2013geometric}. Additionally, since our analysis allows for arbitrary size and wavenumber, we are able to capture the response of the previously unidentified acoustic mode.

\section{Discussion}

While there has been numerical analysis of tessellations in the past, our theoretical formulation provides several key insights into the design and understanding of origami mechanics. We not only analytically calculate expressions for first-order inhomogeneous deformations, but we find an additional acoustic mode of deformation that has not been identified using numerics. Moreover, we have found an analytical expression for a decay length that arises in Miura-ori, and identify that these ``soft modes" are edge states that cannot occur in an infinite tessellation. Indeed, the appearance of a single decay length and the ability to fully quantify the deformation modes using a single wavenumber indicates that the boundaries of Miura-ori fully define the deformation state. We can directly conclude from this that, unlike normal solids, the the number of degrees of freedom scale with the perimeter of a finite tessellation, rather than the area. This result suggests that there are surface boundary states that can be used to probe the full deformation of the material, and hints at the connection between our work at recent studies on topological mechanics \cite{kane2014topological}. In fact, our mathematical formalism shares many parallels with the topological mechanics of linkages \cite{chen2014nonlinear,paulose2015topological,paulose2015selective}, as well as the more conventional literature concerning topological insulators and semimetals \cite{hasan2010colloquium,qi2011topological,po2014phonon}.  It remains to be seen exactly how the symmetry and topology of the crease pattern affect the nature of chiral modes in origami, but there is evidence to suggest that even slight modifications of the crease pattern symmetry may lead to preferentially directed chiral states.

A great deal of this analysis can be carried through to other origami
fold patterns. What is less clear, however, is how the number of degrees of freedom -- the null space of $\mathbf{R}(\mathbf{q})$ -- changes for different fold patterns. At the outset it may seem coincidental that the matrix $\mathbf{R}(\mathbf{q})$ is square. In fact, this behavior is likely more generic. In particular, the Miura-ori -- with additional folds across the faces -- is composed of triangular sub-units. In any triangulated origami fold pattern, vertices will tend to have, on average, six folds. Hence, for $V$ vertices (with $V$ very large), we have $3 V$ unique folds, and $3 V$ degrees of freedom per vertex. Consequently, $\mathbf{R}(\mathbf{q})$ will be a $3V \times 3V$ square matrix for sufficiently large $V$.

Finally, a great advantage to this approach is the ability to separate the topological nature of the crease pattern from the geometry of the vertex. The ability to isolate mechanical deformations or elementary excitations in exotic materials is of great interest in quantum condensed matter \cite{kane2014topological}, amorphous solids \cite{sun2012surface,mao2010soft,wyart2005geometric}, and complex fluids \cite{lerner2012unified}. Our theoretical framework for origami tessellations bridges the gap between the origami mechanics literature and a theory of origami meta-materials by identifying the constraint-based nature of the folding mechanisms and applying well-known methods of analysis from solid state physics and lattice mechanics.

The authors acknowledge interesting and helpful discussions with Tom Hull, Robert Lang, Tomohiro Tachi, Scott Waitukaitis, Martin van Hecke, and Michael Assis. We also thank F. Parish for help with the laser cutter. This work was funded by the National Science Foundation through award EFRI ODISSEI-1240441.

\bibliography{LatticeMechanicsBib}

\begin{thebibliography}{56}%
\makeatletter
\providecommand \@ifxundefined [1]{%
 \@ifx{#1\undefined}
}%
\providecommand \@ifnum [1]{%
 \ifnum #1\expandafter \@firstoftwo
 \else \expandafter \@secondoftwo
 \fi
}%
\providecommand \@ifx [1]{%
 \ifx #1\expandafter \@firstoftwo
 \else \expandafter \@secondoftwo
 \fi
}%
\providecommand \natexlab [1]{#1}%
\providecommand \enquote  [1]{``#1''}%
\providecommand \bibnamefont  [1]{#1}%
\providecommand \bibfnamefont [1]{#1}%
\providecommand \citenamefont [1]{#1}%
\providecommand \href@noop [0]{\@secondoftwo}%
\providecommand \href [0]{\begingroup \@sanitize@url \@href}%
\providecommand \@href[1]{\@@startlink{#1}\@@href}%
\providecommand \@@href[1]{\endgroup#1\@@endlink}%
\providecommand \@sanitize@url [0]{\catcode `\\12\catcode `\$12\catcode
  `\&12\catcode `\#12\catcode `\^12\catcode `\_12\catcode `\%12\relax}%
\providecommand \@@startlink[1]{}%
\providecommand \@@endlink[0]{}%
\providecommand \url  [0]{\begingroup\@sanitize@url \@url }%
\providecommand \@url [1]{\endgroup\@href {#1}{\urlprefix }}%
\providecommand \urlprefix  [0]{URL }%
\providecommand \Eprint [0]{\href }%
\providecommand \doibase [0]{http://dx.doi.org/}%
\providecommand \selectlanguage [0]{\@gobble}%
\providecommand \bibinfo  [0]{\@secondoftwo}%
\providecommand \bibfield  [0]{\@secondoftwo}%
\providecommand \translation [1]{[#1]}%
\providecommand \BibitemOpen [0]{}%
\providecommand \bibitemStop [0]{}%
\providecommand \bibitemNoStop [0]{.\EOS\space}%
\providecommand \EOS [0]{\spacefactor3000\relax}%
\providecommand \BibitemShut  [1]{\csname bibitem#1\endcsname}%
\let\auto@bib@innerbib\@empty
\bibitem [{\citenamefont {Tachi}(2009)}]{tachi2009generalization}%
  \BibitemOpen
  \bibfield  {author} {\bibinfo {author} {\bibfnamefont {T.}~\bibnamefont
  {Tachi}},\ }in\ \href@noop {} {\emph {\bibinfo {booktitle} {Symposium of the
  International Association for Shell and Spatial Structures (50th. 2009.
  Valencia). Evolution and Trends in Design, Analysis and Construction of Shell
  and Spatial Structures: Proceedings}}}\ (\bibinfo {organization} {Editorial
  de la Universitat Polit{\'e}cnica de Valencia.},\ \bibinfo {year}
  {2009})\BibitemShut {NoStop}%
\bibitem [{\citenamefont {Tachi}(2010{\natexlab{a}})}]{tachi2010geometric}%
  \BibitemOpen
  \bibfield  {author} {\bibinfo {author} {\bibfnamefont {T.}~\bibnamefont
  {Tachi}},\ }in\ \href@noop {} {\emph {\bibinfo {booktitle} {Proceedings of
  the International Association for Shell and Spatial Structures (IASS)
  Symposium}}},\ Vol.~\bibinfo {volume} {12}\ (\bibinfo {year} {2010})\ pp.\
  \bibinfo {pages} {458--460}\BibitemShut {NoStop}%
\bibitem [{\citenamefont {Tachi}(2010{\natexlab{b}})}]{tachi2010one}%
  \BibitemOpen
  \bibfield  {author} {\bibinfo {author} {\bibfnamefont {T.}~\bibnamefont
  {Tachi}},\ }in\ \href@noop {} {\emph {\bibinfo {booktitle} {Symposium of the
  International Association for Shell and Spatial Structures (50th. 2009.
  Valencia). Evolution and Trends in Design, Analysis and Construction of Shell
  and Spatial Structures: Proceedings}}}\ (\bibinfo {organization} {Editorial
  de la Universitat Polit{\'e}cnica de Valencia.},\ \bibinfo {year}
  {2010})\BibitemShut {NoStop}%
\bibitem [{\citenamefont {Hawkes}\ \emph {et~al.}(2010)\citenamefont {Hawkes},
  \citenamefont {An}, \citenamefont {Benbernou}, \citenamefont {Tanaka},
  \citenamefont {Kim}, \citenamefont {Demaine}, \citenamefont {Rus},\ and\
  \citenamefont {Wood}}]{hawkes2010programmable}%
  \BibitemOpen
  \bibfield  {author} {\bibinfo {author} {\bibfnamefont {E.}~\bibnamefont
  {Hawkes}}, \bibinfo {author} {\bibfnamefont {B.}~\bibnamefont {An}}, \bibinfo
  {author} {\bibfnamefont {N.}~\bibnamefont {Benbernou}}, \bibinfo {author}
  {\bibfnamefont {H.}~\bibnamefont {Tanaka}}, \bibinfo {author} {\bibfnamefont
  {S.}~\bibnamefont {Kim}}, \bibinfo {author} {\bibfnamefont {E.}~\bibnamefont
  {Demaine}}, \bibinfo {author} {\bibfnamefont {D.}~\bibnamefont {Rus}}, \ and\
  \bibinfo {author} {\bibfnamefont {R.}~\bibnamefont {Wood}},\ }\href@noop {}
  {\bibfield  {journal} {\bibinfo  {journal} {Proc. Natl. Acad. Sci. U.S.A.}\
  }\textbf {\bibinfo {volume} {107}},\ \bibinfo {pages} {12441} (\bibinfo
  {year} {2010})}\BibitemShut {NoStop}%
\bibitem [{\citenamefont {Dias}\ \emph {et~al.}(2012)\citenamefont {Dias},
  \citenamefont {Dudte}, \citenamefont {Mahadevan},\ and\ \citenamefont
  {Santangelo}}]{dias2012geometric}%
  \BibitemOpen
  \bibfield  {author} {\bibinfo {author} {\bibfnamefont {M.~A.}\ \bibnamefont
  {Dias}}, \bibinfo {author} {\bibfnamefont {L.~H.}\ \bibnamefont {Dudte}},
  \bibinfo {author} {\bibfnamefont {L.}~\bibnamefont {Mahadevan}}, \ and\
  \bibinfo {author} {\bibfnamefont {C.~D.}\ \bibnamefont {Santangelo}},\
  }\href@noop {} {\bibfield  {journal} {\bibinfo  {journal} {Phys. Rev. Lett.}\
  }\textbf {\bibinfo {volume} {109}},\ \bibinfo {pages} {114301} (\bibinfo
  {year} {2012})}\BibitemShut {NoStop}%
\bibitem [{\citenamefont {Schenk}\ and\ \citenamefont
  {Guest}(2013)}]{schenk2013geometry}%
  \BibitemOpen
  \bibfield  {author} {\bibinfo {author} {\bibfnamefont {M.}~\bibnamefont
  {Schenk}}\ and\ \bibinfo {author} {\bibfnamefont {S.~D.}\ \bibnamefont
  {Guest}},\ }\href@noop {} {\bibfield  {journal} {\bibinfo  {journal} {Proc.
  Natl. Acad. Sci. U.S.A.}\ }\textbf {\bibinfo {volume} {110}},\ \bibinfo
  {pages} {3276} (\bibinfo {year} {2013})}\BibitemShut {NoStop}%
\bibitem [{\citenamefont {Silverberg}\ \emph {et~al.}(2014)\citenamefont
  {Silverberg}, \citenamefont {Evans}, \citenamefont {McLeod}, \citenamefont
  {Hayward}, \citenamefont {Hull}, \citenamefont {Santangelo},\ and\
  \citenamefont {Cohen}}]{silverberg2014using}%
  \BibitemOpen
  \bibfield  {author} {\bibinfo {author} {\bibfnamefont {J.~L.}\ \bibnamefont
  {Silverberg}}, \bibinfo {author} {\bibfnamefont {A.~A.}\ \bibnamefont
  {Evans}}, \bibinfo {author} {\bibfnamefont {L.}~\bibnamefont {McLeod}},
  \bibinfo {author} {\bibfnamefont {R.~C.}\ \bibnamefont {Hayward}}, \bibinfo
  {author} {\bibfnamefont {T.}~\bibnamefont {Hull}}, \bibinfo {author}
  {\bibfnamefont {C.~D.}\ \bibnamefont {Santangelo}}, \ and\ \bibinfo {author}
  {\bibfnamefont {I.}~\bibnamefont {Cohen}},\ }\href@noop {} {\bibfield
  {journal} {\bibinfo  {journal} {Science}\ }\textbf {\bibinfo {volume}
  {345}},\ \bibinfo {pages} {647} (\bibinfo {year} {2014})}\BibitemShut
  {NoStop}%
\bibitem [{\citenamefont {Na}\ \emph {et~al.}(2014)\citenamefont {Na},
  \citenamefont {Evans}, \citenamefont {Bae}, \citenamefont {Chiappelli},
  \citenamefont {Santangelo}, \citenamefont {Lang}, \citenamefont {Hull},\ and\
  \citenamefont {Hayward}}]{na2014programming}%
  \BibitemOpen
  \bibfield  {author} {\bibinfo {author} {\bibfnamefont {J.-H.}\ \bibnamefont
  {Na}}, \bibinfo {author} {\bibfnamefont {A.~A.}\ \bibnamefont {Evans}},
  \bibinfo {author} {\bibfnamefont {J.}~\bibnamefont {Bae}}, \bibinfo {author}
  {\bibfnamefont {M.~C.}\ \bibnamefont {Chiappelli}}, \bibinfo {author}
  {\bibfnamefont {C.~D.}\ \bibnamefont {Santangelo}}, \bibinfo {author}
  {\bibfnamefont {R.~J.}\ \bibnamefont {Lang}}, \bibinfo {author}
  {\bibfnamefont {T.~C.}\ \bibnamefont {Hull}}, \ and\ \bibinfo {author}
  {\bibfnamefont {R.~C.}\ \bibnamefont {Hayward}},\ }\href@noop {} {\bibfield
  {journal} {\bibinfo  {journal} {Adv. Mater.}\ } (\bibinfo {year}
  {2014})}\BibitemShut {NoStop}%
\bibitem [{\citenamefont {Py}\ \emph {et~al.}(2007)\citenamefont {Py},
  \citenamefont {Reverdy}, \citenamefont {Doppler}, \citenamefont {Bico},
  \citenamefont {Roman},\ and\ \citenamefont {Baroud}}]{py2007capillary}%
  \BibitemOpen
  \bibfield  {author} {\bibinfo {author} {\bibfnamefont {C.}~\bibnamefont
  {Py}}, \bibinfo {author} {\bibfnamefont {P.}~\bibnamefont {Reverdy}},
  \bibinfo {author} {\bibfnamefont {L.}~\bibnamefont {Doppler}}, \bibinfo
  {author} {\bibfnamefont {J.}~\bibnamefont {Bico}}, \bibinfo {author}
  {\bibfnamefont {B.}~\bibnamefont {Roman}}, \ and\ \bibinfo {author}
  {\bibfnamefont {C.~N.}\ \bibnamefont {Baroud}},\ }\href@noop {} {\bibfield
  {journal} {\bibinfo  {journal} {Phys. Rev. Lett.}\ }\textbf {\bibinfo
  {volume} {98}},\ \bibinfo {pages} {156103} (\bibinfo {year}
  {2007})}\BibitemShut {NoStop}%
\bibitem [{\citenamefont {Solomon}\ \emph {et~al.}(2012)\citenamefont
  {Solomon}, \citenamefont {Vouga}, \citenamefont {Wardetzky},\ and\
  \citenamefont {Grinspun}}]{solomon2012flexible}%
  \BibitemOpen
  \bibfield  {author} {\bibinfo {author} {\bibfnamefont {J.}~\bibnamefont
  {Solomon}}, \bibinfo {author} {\bibfnamefont {E.}~\bibnamefont {Vouga}},
  \bibinfo {author} {\bibfnamefont {M.}~\bibnamefont {Wardetzky}}, \ and\
  \bibinfo {author} {\bibfnamefont {E.}~\bibnamefont {Grinspun}},\ }in\
  \href@noop {} {\emph {\bibinfo {booktitle} {Computer Graphics Forum}}},\
  Vol.~\bibinfo {volume} {31}\ (\bibinfo {organization} {Wiley Online
  Library},\ \bibinfo {year} {2012})\ pp.\ \bibinfo {pages}
  {1567--1576}\BibitemShut {NoStop}%
\bibitem [{\citenamefont {Schenk}\ and\ \citenamefont
  {Guest}(2011{\natexlab{a}})}]{schenk2011folded}%
  \BibitemOpen
  \bibfield  {author} {\bibinfo {author} {\bibfnamefont {M.}~\bibnamefont
  {Schenk}}\ and\ \bibinfo {author} {\bibfnamefont {S.}~\bibnamefont {Guest}},\
  }\emph {\bibinfo {title} {Folded shell structures}},\ \href@noop {} {Ph.D.
  thesis},\ \bibinfo  {school} {PhD thesis (Univ of Cambridge, Cambridge,
  United Kingdom)} (\bibinfo {year} {2011}{\natexlab{a}})\BibitemShut {NoStop}%
\bibitem [{\citenamefont {Schenk}\ and\ \citenamefont
  {Guest}(2011{\natexlab{b}})}]{schenk2011origami}%
  \BibitemOpen
  \bibfield  {author} {\bibinfo {author} {\bibfnamefont {M.}~\bibnamefont
  {Schenk}}\ and\ \bibinfo {author} {\bibfnamefont {S.~D.}\ \bibnamefont
  {Guest}},\ }\href@noop {} {\bibfield  {journal} {\bibinfo  {journal}
  {Origami}\ }\textbf {\bibinfo {volume} {5}},\ \bibinfo {pages} {291}
  (\bibinfo {year} {2011}{\natexlab{b}})}\BibitemShut {NoStop}%
\bibitem [{\citenamefont {Wei}\ \emph {et~al.}(2013)\citenamefont {Wei},
  \citenamefont {Guo}, \citenamefont {Dudte}, \citenamefont {Liang},\ and\
  \citenamefont {Mahadevan}}]{wei2013geometric}%
  \BibitemOpen
  \bibfield  {author} {\bibinfo {author} {\bibfnamefont {Z.}~\bibnamefont
  {Wei}}, \bibinfo {author} {\bibfnamefont {Z.}~\bibnamefont {Guo}}, \bibinfo
  {author} {\bibfnamefont {L.}~\bibnamefont {Dudte}}, \bibinfo {author}
  {\bibfnamefont {H.}~\bibnamefont {Liang}}, \ and\ \bibinfo {author}
  {\bibfnamefont {L.}~\bibnamefont {Mahadevan}},\ }\href@noop {} {\bibfield
  {journal} {\bibinfo  {journal} {Phys. Rev. Lett.}\ }\textbf {\bibinfo
  {volume} {110}},\ \bibinfo {pages} {215501} (\bibinfo {year}
  {2013})}\BibitemShut {NoStop}%
\bibitem [{\citenamefont {Abdul-Sater}\ \emph {et~al.}(2013)\citenamefont
  {Abdul-Sater}, \citenamefont {Irlinger},\ and\ \citenamefont
  {Lueth}}]{abdul2013two}%
  \BibitemOpen
  \bibfield  {author} {\bibinfo {author} {\bibfnamefont {K.}~\bibnamefont
  {Abdul-Sater}}, \bibinfo {author} {\bibfnamefont {F.}~\bibnamefont
  {Irlinger}}, \ and\ \bibinfo {author} {\bibfnamefont {T.~C.}\ \bibnamefont
  {Lueth}},\ }\href@noop {} {\bibfield  {journal} {\bibinfo  {journal} {J.
  Mech. Robot.}\ }\textbf {\bibinfo {volume} {5}},\ \bibinfo {pages} {031005}
  (\bibinfo {year} {2013})}\BibitemShut {NoStop}%
\bibitem [{\citenamefont {Waitukaitis}\ \emph {et~al.}(2015)\citenamefont
  {Waitukaitis}, \citenamefont {Menaut}, \citenamefont {Chen},\ and\
  \citenamefont {van Hecke}}]{waitukaitis2015origami}%
  \BibitemOpen
  \bibfield  {author} {\bibinfo {author} {\bibfnamefont {S.}~\bibnamefont
  {Waitukaitis}}, \bibinfo {author} {\bibfnamefont {R.}~\bibnamefont {Menaut}},
  \bibinfo {author} {\bibfnamefont {B.~G.-g.}\ \bibnamefont {Chen}}, \ and\
  \bibinfo {author} {\bibfnamefont {M.}~\bibnamefont {van Hecke}},\ }\href@noop
  {} {\bibfield  {journal} {\bibinfo  {journal} {Phys. Rev. Lett.}\ }\textbf
  {\bibinfo {volume} {114}},\ \bibinfo {pages} {055503} (\bibinfo {year}
  {2015})}\BibitemShut {NoStop}%
\bibitem [{\citenamefont {Hanna}\ \emph {et~al.}(2014)\citenamefont {Hanna},
  \citenamefont {Lund}, \citenamefont {Lang}, \citenamefont {Magleby},\ and\
  \citenamefont {Howell}}]{hanna2014waterbomb}%
  \BibitemOpen
  \bibfield  {author} {\bibinfo {author} {\bibfnamefont {B.~H.}\ \bibnamefont
  {Hanna}}, \bibinfo {author} {\bibfnamefont {J.~M.}\ \bibnamefont {Lund}},
  \bibinfo {author} {\bibfnamefont {R.~J.}\ \bibnamefont {Lang}}, \bibinfo
  {author} {\bibfnamefont {S.~P.}\ \bibnamefont {Magleby}}, \ and\ \bibinfo
  {author} {\bibfnamefont {L.~L.}\ \bibnamefont {Howell}},\ }\href@noop {}
  {\bibfield  {journal} {\bibinfo  {journal} {Smart Mater. Struct.}\ }\textbf
  {\bibinfo {volume} {23}},\ \bibinfo {pages} {094009} (\bibinfo {year}
  {2014})}\BibitemShut {NoStop}%
\bibitem [{\citenamefont {Bende}\ \emph {et~al.}(2014)\citenamefont {Bende},
  \citenamefont {Evans}, \citenamefont {Innes-Gold}, \citenamefont {Marin},
  \citenamefont {Cohen}, \citenamefont {Hayward},\ and\ \citenamefont
  {Santangelo}}]{bende2014geometrically}%
  \BibitemOpen
  \bibfield  {author} {\bibinfo {author} {\bibfnamefont {N.~P.}\ \bibnamefont
  {Bende}}, \bibinfo {author} {\bibfnamefont {A.~A.}\ \bibnamefont {Evans}},
  \bibinfo {author} {\bibfnamefont {S.}~\bibnamefont {Innes-Gold}}, \bibinfo
  {author} {\bibfnamefont {L.~A.}\ \bibnamefont {Marin}}, \bibinfo {author}
  {\bibfnamefont {I.}~\bibnamefont {Cohen}}, \bibinfo {author} {\bibfnamefont
  {R.~C.}\ \bibnamefont {Hayward}}, \ and\ \bibinfo {author} {\bibfnamefont
  {C.~D.}\ \bibnamefont {Santangelo}},\ }\href@noop {} {\bibfield  {journal}
  {\bibinfo  {journal} {arXiv preprint arXiv:1410.7038}\ } (\bibinfo {year}
  {2014})}\BibitemShut {NoStop}%
\bibitem [{\citenamefont {Pendry}\ \emph {et~al.}(2006)\citenamefont {Pendry},
  \citenamefont {Schurig},\ and\ \citenamefont
  {Smith}}]{pendry2006controlling}%
  \BibitemOpen
  \bibfield  {author} {\bibinfo {author} {\bibfnamefont {J.~B.}\ \bibnamefont
  {Pendry}}, \bibinfo {author} {\bibfnamefont {D.}~\bibnamefont {Schurig}}, \
  and\ \bibinfo {author} {\bibfnamefont {D.~R.}\ \bibnamefont {Smith}},\
  }\href@noop {} {\bibfield  {journal} {\bibinfo  {journal} {Science}\ }\textbf
  {\bibinfo {volume} {312}},\ \bibinfo {pages} {1780} (\bibinfo {year}
  {2006})}\BibitemShut {NoStop}%
\bibitem [{\citenamefont {Kadic}\ \emph {et~al.}(2012)\citenamefont {Kadic},
  \citenamefont {B{\"u}ckmann}, \citenamefont {Stenger}, \citenamefont
  {Thiel},\ and\ \citenamefont {Wegener}}]{kadic2012practicability}%
  \BibitemOpen
  \bibfield  {author} {\bibinfo {author} {\bibfnamefont {M.}~\bibnamefont
  {Kadic}}, \bibinfo {author} {\bibfnamefont {T.}~\bibnamefont {B{\"u}ckmann}},
  \bibinfo {author} {\bibfnamefont {N.}~\bibnamefont {Stenger}}, \bibinfo
  {author} {\bibfnamefont {M.}~\bibnamefont {Thiel}}, \ and\ \bibinfo {author}
  {\bibfnamefont {M.}~\bibnamefont {Wegener}},\ }\href@noop {} {\bibfield
  {journal} {\bibinfo  {journal} {Appl. Phys. Lett.}\ }\textbf {\bibinfo
  {volume} {100}},\ \bibinfo {pages} {191901} (\bibinfo {year}
  {2012})}\BibitemShut {NoStop}%
\bibitem [{\citenamefont {Brule}\ \emph {et~al.}(2014)\citenamefont {Brule},
  \citenamefont {Javelaud}, \citenamefont {Enoch},\ and\ \citenamefont
  {Guenneau}}]{seismic2014}%
  \BibitemOpen
  \bibfield  {author} {\bibinfo {author} {\bibfnamefont {S.}~\bibnamefont
  {Brule}}, \bibinfo {author} {\bibfnamefont {E.}~\bibnamefont {Javelaud}},
  \bibinfo {author} {\bibfnamefont {S.}~\bibnamefont {Enoch}}, \ and\ \bibinfo
  {author} {\bibfnamefont {S.}~\bibnamefont {Guenneau}},\ }\href@noop {}
  {\bibfield  {journal} {\bibinfo  {journal} {Phys. Rev. Lett.}\ }\textbf
  {\bibinfo {volume} {112}},\ \bibinfo {pages} {133901} (\bibinfo {year}
  {2014})}\BibitemShut {NoStop}%
\bibitem [{\citenamefont {Farhat}\ \emph {et~al.}(2009)\citenamefont {Farhat},
  \citenamefont {Guenneau},\ and\ \citenamefont
  {Enoch}}]{farhat2009ultrabroadband}%
  \BibitemOpen
  \bibfield  {author} {\bibinfo {author} {\bibfnamefont {M.}~\bibnamefont
  {Farhat}}, \bibinfo {author} {\bibfnamefont {S.}~\bibnamefont {Guenneau}}, \
  and\ \bibinfo {author} {\bibfnamefont {S.}~\bibnamefont {Enoch}},\
  }\href@noop {} {\bibfield  {journal} {\bibinfo  {journal} {Phys. Rev. Lett.}\
  }\textbf {\bibinfo {volume} {103}},\ \bibinfo {pages} {024301} (\bibinfo
  {year} {2009})}\BibitemShut {NoStop}%
\bibitem [{\citenamefont {Stenger}\ \emph {et~al.}(2012)\citenamefont
  {Stenger}, \citenamefont {Wilhelm},\ and\ \citenamefont
  {Wegener}}]{stenger2012experiments}%
  \BibitemOpen
  \bibfield  {author} {\bibinfo {author} {\bibfnamefont {N.}~\bibnamefont
  {Stenger}}, \bibinfo {author} {\bibfnamefont {M.}~\bibnamefont {Wilhelm}}, \
  and\ \bibinfo {author} {\bibfnamefont {M.}~\bibnamefont {Wegener}},\
  }\href@noop {} {\bibfield  {journal} {\bibinfo  {journal} {Phys. Rev. Lett.}\
  }\textbf {\bibinfo {volume} {108}},\ \bibinfo {pages} {014301} (\bibinfo
  {year} {2012})}\BibitemShut {NoStop}%
\bibitem [{\citenamefont {Shim}\ \emph {et~al.}(2013)\citenamefont {Shim},
  \citenamefont {Shan}, \citenamefont {Ko{\v{s}}mrlj}, \citenamefont {Kang},
  \citenamefont {Chen}, \citenamefont {Weaver},\ and\ \citenamefont
  {Bertoldi}}]{shim2013harnessing}%
  \BibitemOpen
  \bibfield  {author} {\bibinfo {author} {\bibfnamefont {J.}~\bibnamefont
  {Shim}}, \bibinfo {author} {\bibfnamefont {S.}~\bibnamefont {Shan}}, \bibinfo
  {author} {\bibfnamefont {A.}~\bibnamefont {Ko{\v{s}}mrlj}}, \bibinfo {author}
  {\bibfnamefont {S.~H.}\ \bibnamefont {Kang}}, \bibinfo {author}
  {\bibfnamefont {E.~R.}\ \bibnamefont {Chen}}, \bibinfo {author}
  {\bibfnamefont {J.~C.}\ \bibnamefont {Weaver}}, \ and\ \bibinfo {author}
  {\bibfnamefont {K.}~\bibnamefont {Bertoldi}},\ }\href@noop {} {\bibfield
  {journal} {\bibinfo  {journal} {Soft Matter}\ }\textbf {\bibinfo {volume}
  {9}},\ \bibinfo {pages} {8198} (\bibinfo {year} {2013})}\BibitemShut
  {NoStop}%
\bibitem [{\citenamefont {Evans}\ and\ \citenamefont
  {Levine}(2013)}]{evans2013reflection}%
  \BibitemOpen
  \bibfield  {author} {\bibinfo {author} {\bibfnamefont {A.~A.}\ \bibnamefont
  {Evans}}\ and\ \bibinfo {author} {\bibfnamefont {A.~J.}\ \bibnamefont
  {Levine}},\ }\href@noop {} {\bibfield  {journal} {\bibinfo  {journal} {Phys.
  Rev. Lett.}\ }\textbf {\bibinfo {volume} {111}},\ \bibinfo {pages} {038101}
  (\bibinfo {year} {2013})}\BibitemShut {NoStop}%
\bibitem [{\citenamefont {Zhang}\ \emph {et~al.}(2008)\citenamefont {Zhang},
  \citenamefont {Matsumoto}, \citenamefont {Peter}, \citenamefont {Lin},
  \citenamefont {Kamien},\ and\ \citenamefont {Yang}}]{zhang2008one}%
  \BibitemOpen
  \bibfield  {author} {\bibinfo {author} {\bibfnamefont {Y.}~\bibnamefont
  {Zhang}}, \bibinfo {author} {\bibfnamefont {E.~A.}\ \bibnamefont
  {Matsumoto}}, \bibinfo {author} {\bibfnamefont {A.}~\bibnamefont {Peter}},
  \bibinfo {author} {\bibfnamefont {P.-C.}\ \bibnamefont {Lin}}, \bibinfo
  {author} {\bibfnamefont {R.~D.}\ \bibnamefont {Kamien}}, \ and\ \bibinfo
  {author} {\bibfnamefont {S.}~\bibnamefont {Yang}},\ }\href@noop {} {\bibfield
   {journal} {\bibinfo  {journal} {Nano Lett.}\ }\textbf {\bibinfo {volume}
  {8}},\ \bibinfo {pages} {1192} (\bibinfo {year} {2008})}\BibitemShut
  {NoStop}%
\bibitem [{\citenamefont {Matsumoto}\ and\ \citenamefont
  {Kamien}(2009)}]{matsumoto2009elastic}%
  \BibitemOpen
  \bibfield  {author} {\bibinfo {author} {\bibfnamefont {E.~A.}\ \bibnamefont
  {Matsumoto}}\ and\ \bibinfo {author} {\bibfnamefont {R.~D.}\ \bibnamefont
  {Kamien}},\ }\href@noop {} {\bibfield  {journal} {\bibinfo  {journal} {Phys.
  Rev. E}\ }\textbf {\bibinfo {volume} {80}},\ \bibinfo {pages} {021604}
  (\bibinfo {year} {2009})}\BibitemShut {NoStop}%
\bibitem [{\citenamefont {Bertoldi}\ \emph {et~al.}(2010)\citenamefont
  {Bertoldi}, \citenamefont {Reis}, \citenamefont {Willshaw},\ and\
  \citenamefont {Mullin}}]{bertoldi2010negative}%
  \BibitemOpen
  \bibfield  {author} {\bibinfo {author} {\bibfnamefont {K.}~\bibnamefont
  {Bertoldi}}, \bibinfo {author} {\bibfnamefont {P.~M.}\ \bibnamefont {Reis}},
  \bibinfo {author} {\bibfnamefont {S.}~\bibnamefont {Willshaw}}, \ and\
  \bibinfo {author} {\bibfnamefont {T.}~\bibnamefont {Mullin}},\ }\href@noop {}
  {\bibfield  {journal} {\bibinfo  {journal} {Adv. Mater.}\ }\textbf {\bibinfo
  {volume} {22}},\ \bibinfo {pages} {361} (\bibinfo {year} {2010})}\BibitemShut
  {NoStop}%
\bibitem [{\citenamefont {Matsumoto}\ and\ \citenamefont
  {Kamien}(2012)}]{matsumoto2012patterns}%
  \BibitemOpen
  \bibfield  {author} {\bibinfo {author} {\bibfnamefont {E.~A.}\ \bibnamefont
  {Matsumoto}}\ and\ \bibinfo {author} {\bibfnamefont {R.~D.}\ \bibnamefont
  {Kamien}},\ }\href@noop {} {\bibfield  {journal} {\bibinfo  {journal} {Soft
  Matter}\ }\textbf {\bibinfo {volume} {8}},\ \bibinfo {pages} {11038}
  (\bibinfo {year} {2012})}\BibitemShut {NoStop}%
\bibitem [{\citenamefont {Overvelde}\ \emph {et~al.}(2012)\citenamefont
  {Overvelde}, \citenamefont {Shan},\ and\ \citenamefont
  {Bertoldi}}]{overvelde2012compaction}%
  \BibitemOpen
  \bibfield  {author} {\bibinfo {author} {\bibfnamefont {J.~T.~B.}\
  \bibnamefont {Overvelde}}, \bibinfo {author} {\bibfnamefont {S.}~\bibnamefont
  {Shan}}, \ and\ \bibinfo {author} {\bibfnamefont {K.}~\bibnamefont
  {Bertoldi}},\ }\href@noop {} {\bibfield  {journal} {\bibinfo  {journal} {Adv.
  Mater.}\ }\textbf {\bibinfo {volume} {24}},\ \bibinfo {pages} {2337}
  (\bibinfo {year} {2012})}\BibitemShut {NoStop}%
\bibitem [{tes()}]{tessellatica}%
  \BibitemOpen
  \href@noop {} {\enquote {\bibinfo {title} {Tessellatica},}\ }\bibinfo
  {howpublished}
  {\url{http://www.langorigami.com/science/computational/tessellatica/tessellatica.php}}\BibitemShut
  {NoStop}%
\bibitem [{\citenamefont {Yoon}\ \emph {et~al.}(2014)\citenamefont {Yoon},
  \citenamefont {Xiao}, \citenamefont {Park}, \citenamefont {Cha},
  \citenamefont {Nguyen},\ and\ \citenamefont {Gracias}}]{yoon2014functional}%
  \BibitemOpen
  \bibfield  {author} {\bibinfo {author} {\bibfnamefont {C.}~\bibnamefont
  {Yoon}}, \bibinfo {author} {\bibfnamefont {R.}~\bibnamefont {Xiao}}, \bibinfo
  {author} {\bibfnamefont {J.}~\bibnamefont {Park}}, \bibinfo {author}
  {\bibfnamefont {J.}~\bibnamefont {Cha}}, \bibinfo {author} {\bibfnamefont
  {T.~D.}\ \bibnamefont {Nguyen}}, \ and\ \bibinfo {author} {\bibfnamefont
  {D.~H.}\ \bibnamefont {Gracias}},\ }\href@noop {} {\bibfield  {journal}
  {\bibinfo  {journal} {Smart Mater. Struct.}\ }\textbf {\bibinfo {volume}
  {23}},\ \bibinfo {pages} {094008} (\bibinfo {year} {2014})}\BibitemShut
  {NoStop}%
\bibitem [{\citenamefont {Liu}\ \emph {et~al.}(2012)\citenamefont {Liu},
  \citenamefont {Boyles}, \citenamefont {Genzer},\ and\ \citenamefont
  {Dickey}}]{liu2012self}%
  \BibitemOpen
  \bibfield  {author} {\bibinfo {author} {\bibfnamefont {Y.}~\bibnamefont
  {Liu}}, \bibinfo {author} {\bibfnamefont {J.~K.}\ \bibnamefont {Boyles}},
  \bibinfo {author} {\bibfnamefont {J.}~\bibnamefont {Genzer}}, \ and\ \bibinfo
  {author} {\bibfnamefont {M.~D.}\ \bibnamefont {Dickey}},\ }\href@noop {}
  {\bibfield  {journal} {\bibinfo  {journal} {Soft Matter}\ }\textbf {\bibinfo
  {volume} {8}},\ \bibinfo {pages} {1764} (\bibinfo {year} {2012})}\BibitemShut
  {NoStop}%
\bibitem [{\citenamefont {Ionov}(2011)}]{ionov2011soft}%
  \BibitemOpen
  \bibfield  {author} {\bibinfo {author} {\bibfnamefont {L.}~\bibnamefont
  {Ionov}},\ }\href@noop {} {\bibfield  {journal} {\bibinfo  {journal} {Soft
  Matter}\ }\textbf {\bibinfo {volume} {7}},\ \bibinfo {pages} {6786} (\bibinfo
  {year} {2011})}\BibitemShut {NoStop}%
\bibitem [{\citenamefont {Stoychev}\ \emph {et~al.}(2011)\citenamefont
  {Stoychev}, \citenamefont {Puretskiy},\ and\ \citenamefont
  {Ionov}}]{stoychev2011self}%
  \BibitemOpen
  \bibfield  {author} {\bibinfo {author} {\bibfnamefont {G.}~\bibnamefont
  {Stoychev}}, \bibinfo {author} {\bibfnamefont {N.}~\bibnamefont {Puretskiy}},
  \ and\ \bibinfo {author} {\bibfnamefont {L.}~\bibnamefont {Ionov}},\
  }\href@noop {} {\bibfield  {journal} {\bibinfo  {journal} {Soft Matter}\
  }\textbf {\bibinfo {volume} {7}},\ \bibinfo {pages} {3277} (\bibinfo {year}
  {2011})}\BibitemShut {NoStop}%
\bibitem [{\citenamefont {Huffman}(1976)}]{huffman1976curvature}%
  \BibitemOpen
  \bibfield  {author} {\bibinfo {author} {\bibfnamefont {D.~A.}\ \bibnamefont
  {Huffman}},\ }\href@noop {} {\bibfield  {journal} {\bibinfo  {journal} {IEEE
  Trans. Computers}\ }\textbf {\bibinfo {volume} {25}},\ \bibinfo {pages}
  {1010} (\bibinfo {year} {1976})}\BibitemShut {NoStop}%
\bibitem [{\citenamefont {Hull}\ and\ \citenamefont
  {belcastro}(2002)}]{hull2002modelling}%
  \BibitemOpen
  \bibfield  {author} {\bibinfo {author} {\bibfnamefont {T.~C.}\ \bibnamefont
  {Hull}}\ and\ \bibinfo {author} {\bibfnamefont {s.-m.}\ \bibnamefont
  {belcastro}},\ }\href@noop {} {\bibfield  {journal} {\bibinfo  {journal}
  {Linear Algebra Appl.}\ }\textbf {\bibinfo {volume} {348}},\ \bibinfo {pages}
  {273} (\bibinfo {year} {2002})}\BibitemShut {NoStop}%
\bibitem [{\citenamefont {Hutchinson}\ and\ \citenamefont
  {Fleck}(2006)}]{hutchinson2006structural}%
  \BibitemOpen
  \bibfield  {author} {\bibinfo {author} {\bibfnamefont {R.}~\bibnamefont
  {Hutchinson}}\ and\ \bibinfo {author} {\bibfnamefont {N.}~\bibnamefont
  {Fleck}},\ }\href@noop {} {\bibfield  {journal} {\bibinfo  {journal} {J.
  Mech. Phys. Solids}\ }\textbf {\bibinfo {volume} {54}},\ \bibinfo {pages}
  {756} (\bibinfo {year} {2006})}\BibitemShut {NoStop}%
\bibitem [{\citenamefont {Kane}\ and\ \citenamefont
  {Lubensky}(2014)}]{kane2014topological}%
  \BibitemOpen
  \bibfield  {author} {\bibinfo {author} {\bibfnamefont {C.}~\bibnamefont
  {Kane}}\ and\ \bibinfo {author} {\bibfnamefont {T.}~\bibnamefont
  {Lubensky}},\ }\href@noop {} {\bibfield  {journal} {\bibinfo  {journal}
  {Nature Phys.}\ }\textbf {\bibinfo {volume} {10}},\ \bibinfo {pages} {39}
  (\bibinfo {year} {2014})}\BibitemShut {NoStop}%
\bibitem [{\citenamefont {Lechenault}\ \emph {et~al.}(2014)\citenamefont
  {Lechenault}, \citenamefont {Thiria},\ and\ \citenamefont
  {Adda-Bedia}}]{lechenault2014mechanical}%
  \BibitemOpen
  \bibfield  {author} {\bibinfo {author} {\bibfnamefont {F.}~\bibnamefont
  {Lechenault}}, \bibinfo {author} {\bibfnamefont {B.}~\bibnamefont {Thiria}},
  \ and\ \bibinfo {author} {\bibfnamefont {M.}~\bibnamefont {Adda-Bedia}},\
  }\href@noop {} {\bibfield  {journal} {\bibinfo  {journal} {Phys. Rev. Lett.}\
  }\textbf {\bibinfo {volume} {112}},\ \bibinfo {pages} {244301} (\bibinfo
  {year} {2014})}\BibitemShut {NoStop}%
\bibitem [{\citenamefont {Silverberg}\ \emph {et~al.}(2015)\citenamefont
  {Silverberg}, \citenamefont {Na}, \citenamefont {Evans}, \citenamefont {Liu},
  \citenamefont {Hull}, \citenamefont {Santangelo}, \citenamefont {Lang},
  \citenamefont {Hayward},\ and\ \citenamefont
  {Cohen}}]{silverberg2015origami}%
  \BibitemOpen
  \bibfield  {author} {\bibinfo {author} {\bibfnamefont {J.~L.}\ \bibnamefont
  {Silverberg}}, \bibinfo {author} {\bibfnamefont {J.-H.}\ \bibnamefont {Na}},
  \bibinfo {author} {\bibfnamefont {A.~A.}\ \bibnamefont {Evans}}, \bibinfo
  {author} {\bibfnamefont {B.}~\bibnamefont {Liu}}, \bibinfo {author}
  {\bibfnamefont {T.~C.}\ \bibnamefont {Hull}}, \bibinfo {author}
  {\bibfnamefont {C.~D.}\ \bibnamefont {Santangelo}}, \bibinfo {author}
  {\bibfnamefont {R.~J.}\ \bibnamefont {Lang}}, \bibinfo {author}
  {\bibfnamefont {R.~C.}\ \bibnamefont {Hayward}}, \ and\ \bibinfo {author}
  {\bibfnamefont {I.}~\bibnamefont {Cohen}},\ }\href@noop {} {\bibfield
  {journal} {\bibinfo  {journal} {Nature materials}\ }\textbf {\bibinfo
  {volume} {14}},\ \bibinfo {pages} {389} (\bibinfo {year} {2015})}\BibitemShut
  {NoStop}%
\bibitem [{\citenamefont {Mahadevan}\ and\ \citenamefont
  {Rica}(2005)}]{mahadevan2005self}%
  \BibitemOpen
  \bibfield  {author} {\bibinfo {author} {\bibfnamefont {L.}~\bibnamefont
  {Mahadevan}}\ and\ \bibinfo {author} {\bibfnamefont {S.}~\bibnamefont
  {Rica}},\ }\href@noop {} {\bibfield  {journal} {\bibinfo  {journal}
  {Science}\ }\textbf {\bibinfo {volume} {307}},\ \bibinfo {pages} {1740}
  (\bibinfo {year} {2005})}\BibitemShut {NoStop}%
\bibitem [{\citenamefont {Shyer}\ \emph {et~al.}(2013)\citenamefont {Shyer},
  \citenamefont {Tallinen}, \citenamefont {Nerurkar}, \citenamefont {Wei},
  \citenamefont {Gil}, \citenamefont {Kaplan}, \citenamefont {Tabin},\ and\
  \citenamefont {Mahadevan}}]{shyer2013villification}%
  \BibitemOpen
  \bibfield  {author} {\bibinfo {author} {\bibfnamefont {A.~E.}\ \bibnamefont
  {Shyer}}, \bibinfo {author} {\bibfnamefont {T.}~\bibnamefont {Tallinen}},
  \bibinfo {author} {\bibfnamefont {N.~L.}\ \bibnamefont {Nerurkar}}, \bibinfo
  {author} {\bibfnamefont {Z.}~\bibnamefont {Wei}}, \bibinfo {author}
  {\bibfnamefont {E.~S.}\ \bibnamefont {Gil}}, \bibinfo {author} {\bibfnamefont
  {D.~L.}\ \bibnamefont {Kaplan}}, \bibinfo {author} {\bibfnamefont {C.~J.}\
  \bibnamefont {Tabin}}, \ and\ \bibinfo {author} {\bibfnamefont
  {L.}~\bibnamefont {Mahadevan}},\ }\href@noop {} {\bibfield  {journal}
  {\bibinfo  {journal} {Science}\ }\textbf {\bibinfo {volume} {342}},\ \bibinfo
  {pages} {212} (\bibinfo {year} {2013})}\BibitemShut {NoStop}%
\bibitem [{\citenamefont {Demaine}\ \emph {et~al.}(2011)\citenamefont
  {Demaine}, \citenamefont {Demaine}, \citenamefont {Hart}, \citenamefont
  {Price},\ and\ \citenamefont {Tachi}}]{demaine2011non}%
  \BibitemOpen
  \bibfield  {author} {\bibinfo {author} {\bibfnamefont {E.~D.}\ \bibnamefont
  {Demaine}}, \bibinfo {author} {\bibfnamefont {M.~L.}\ \bibnamefont
  {Demaine}}, \bibinfo {author} {\bibfnamefont {V.}~\bibnamefont {Hart}},
  \bibinfo {author} {\bibfnamefont {G.~N.}\ \bibnamefont {Price}}, \ and\
  \bibinfo {author} {\bibfnamefont {T.}~\bibnamefont {Tachi}},\ }\href@noop {}
  {\bibfield  {journal} {\bibinfo  {journal} {Graphs and Combinatorics}\
  }\textbf {\bibinfo {volume} {27}},\ \bibinfo {pages} {377} (\bibinfo {year}
  {2011})}\BibitemShut {NoStop}%
\bibitem [{\citenamefont {Witten}(2007)}]{witten2007stress}%
  \BibitemOpen
  \bibfield  {author} {\bibinfo {author} {\bibfnamefont {T.}~\bibnamefont
  {Witten}},\ }\href@noop {} {\bibfield  {journal} {\bibinfo  {journal} {Rev.
  Mod. Phys.}\ }\textbf {\bibinfo {volume} {79}},\ \bibinfo {pages} {643}
  (\bibinfo {year} {2007})}\BibitemShut {NoStop}%
\bibitem [{\citenamefont {Phani}\ and\ \citenamefont
  {Fleck}(2008)}]{phani2008elastic}%
  \BibitemOpen
  \bibfield  {author} {\bibinfo {author} {\bibfnamefont {A.~S.}\ \bibnamefont
  {Phani}}\ and\ \bibinfo {author} {\bibfnamefont {N.~A.}\ \bibnamefont
  {Fleck}},\ }\href@noop {} {\bibfield  {journal} {\bibinfo  {journal} {J.
  Appl. Mech.}\ }\textbf {\bibinfo {volume} {75}},\ \bibinfo {pages} {021020}
  (\bibinfo {year} {2008})}\BibitemShut {NoStop}%
\bibitem [{\citenamefont {Strutt}\ and\ \citenamefont
  {Rayleigh}(1885)}]{strutt1885waves}%
  \BibitemOpen
  \bibfield  {author} {\bibinfo {author} {\bibfnamefont {J.~W.}\ \bibnamefont
  {Strutt}}\ and\ \bibinfo {author} {\bibfnamefont {L.}~\bibnamefont
  {Rayleigh}},\ }\href@noop {} {\bibfield  {journal} {\bibinfo  {journal}
  {Proceedings of the London Mathematical Society}\ }\textbf {\bibinfo {volume}
  {17}},\ \bibinfo {pages} {4} (\bibinfo {year} {1885})}\BibitemShut {NoStop}%
\bibitem [{\citenamefont {Chen}\ \emph {et~al.}(2014)\citenamefont {Chen},
  \citenamefont {Upadhyaya},\ and\ \citenamefont
  {Vitelli}}]{chen2014nonlinear}%
  \BibitemOpen
  \bibfield  {author} {\bibinfo {author} {\bibfnamefont {B.~G.-g.}\
  \bibnamefont {Chen}}, \bibinfo {author} {\bibfnamefont {N.}~\bibnamefont
  {Upadhyaya}}, \ and\ \bibinfo {author} {\bibfnamefont {V.}~\bibnamefont
  {Vitelli}},\ }\href@noop {} {\bibfield  {journal} {\bibinfo  {journal}
  {Proceedings of the National Academy of Sciences}\ }\textbf {\bibinfo
  {volume} {111}},\ \bibinfo {pages} {13004} (\bibinfo {year}
  {2014})}\BibitemShut {NoStop}%
\bibitem [{\citenamefont {Paulose}\ \emph
  {et~al.}(2015{\natexlab{a}})\citenamefont {Paulose}, \citenamefont {Chen},\
  and\ \citenamefont {Vitelli}}]{paulose2015topological}%
  \BibitemOpen
  \bibfield  {author} {\bibinfo {author} {\bibfnamefont {J.}~\bibnamefont
  {Paulose}}, \bibinfo {author} {\bibfnamefont {B.~G.-g.}\ \bibnamefont
  {Chen}}, \ and\ \bibinfo {author} {\bibfnamefont {V.}~\bibnamefont
  {Vitelli}},\ }\href@noop {} {\bibfield  {journal} {\bibinfo  {journal}
  {Nature Phys.}\ } (\bibinfo {year} {2015}{\natexlab{a}})}\BibitemShut
  {NoStop}%
\bibitem [{\citenamefont {Paulose}\ \emph
  {et~al.}(2015{\natexlab{b}})\citenamefont {Paulose}, \citenamefont
  {Meeussen},\ and\ \citenamefont {Vitelli}}]{paulose2015selective}%
  \BibitemOpen
  \bibfield  {author} {\bibinfo {author} {\bibfnamefont {J.}~\bibnamefont
  {Paulose}}, \bibinfo {author} {\bibfnamefont {A.~S.}\ \bibnamefont
  {Meeussen}}, \ and\ \bibinfo {author} {\bibfnamefont {V.}~\bibnamefont
  {Vitelli}},\ }\href@noop {} {\bibfield  {journal} {\bibinfo  {journal} {arXiv
  preprint arXiv:1502.03396}\ } (\bibinfo {year}
  {2015}{\natexlab{b}})}\BibitemShut {NoStop}%
\bibitem [{\citenamefont {Hasan}\ and\ \citenamefont
  {Kane}(2010)}]{hasan2010colloquium}%
  \BibitemOpen
  \bibfield  {author} {\bibinfo {author} {\bibfnamefont {M.~Z.}\ \bibnamefont
  {Hasan}}\ and\ \bibinfo {author} {\bibfnamefont {C.~L.}\ \bibnamefont
  {Kane}},\ }\href@noop {} {\bibfield  {journal} {\bibinfo  {journal} {Rev.
  Mod. Phys.}\ }\textbf {\bibinfo {volume} {82}},\ \bibinfo {pages} {3045}
  (\bibinfo {year} {2010})}\BibitemShut {NoStop}%
\bibitem [{\citenamefont {Qi}\ and\ \citenamefont
  {Zhang}(2011)}]{qi2011topological}%
  \BibitemOpen
  \bibfield  {author} {\bibinfo {author} {\bibfnamefont {X.-L.}\ \bibnamefont
  {Qi}}\ and\ \bibinfo {author} {\bibfnamefont {S.-C.}\ \bibnamefont {Zhang}},\
  }\href@noop {} {\bibfield  {journal} {\bibinfo  {journal} {Rev. Mod. Phys.}\
  }\textbf {\bibinfo {volume} {83}},\ \bibinfo {pages} {1057} (\bibinfo {year}
  {2011})}\BibitemShut {NoStop}%
\bibitem [{\citenamefont {Po}\ \emph {et~al.}(2014)\citenamefont {Po},
  \citenamefont {Bahri},\ and\ \citenamefont {Vishwanath}}]{po2014phonon}%
  \BibitemOpen
  \bibfield  {author} {\bibinfo {author} {\bibfnamefont {H.~C.}\ \bibnamefont
  {Po}}, \bibinfo {author} {\bibfnamefont {Y.}~\bibnamefont {Bahri}}, \ and\
  \bibinfo {author} {\bibfnamefont {A.}~\bibnamefont {Vishwanath}},\
  }\href@noop {} {\bibfield  {journal} {\bibinfo  {journal} {arXiv preprint
  arXiv:1410.1320}\ } (\bibinfo {year} {2014})}\BibitemShut {NoStop}%
\bibitem [{\citenamefont {Sun}\ \emph {et~al.}(2012)\citenamefont {Sun},
  \citenamefont {Souslov}, \citenamefont {Mao},\ and\ \citenamefont
  {Lubensky}}]{sun2012surface}%
  \BibitemOpen
  \bibfield  {author} {\bibinfo {author} {\bibfnamefont {K.}~\bibnamefont
  {Sun}}, \bibinfo {author} {\bibfnamefont {A.}~\bibnamefont {Souslov}},
  \bibinfo {author} {\bibfnamefont {X.}~\bibnamefont {Mao}}, \ and\ \bibinfo
  {author} {\bibfnamefont {T.}~\bibnamefont {Lubensky}},\ }\href@noop {}
  {\bibfield  {journal} {\bibinfo  {journal} {Proc. Natl. Acad. Sci. U.S.A.}\
  }\textbf {\bibinfo {volume} {109}},\ \bibinfo {pages} {12369} (\bibinfo
  {year} {2012})}\BibitemShut {NoStop}%
\bibitem [{\citenamefont {Mao}\ \emph {et~al.}(2010)\citenamefont {Mao},
  \citenamefont {Xu},\ and\ \citenamefont {Lubensky}}]{mao2010soft}%
  \BibitemOpen
  \bibfield  {author} {\bibinfo {author} {\bibfnamefont {X.}~\bibnamefont
  {Mao}}, \bibinfo {author} {\bibfnamefont {N.}~\bibnamefont {Xu}}, \ and\
  \bibinfo {author} {\bibfnamefont {T.}~\bibnamefont {Lubensky}},\ }\href@noop
  {} {\bibfield  {journal} {\bibinfo  {journal} {Phys. Rev. Lett.}\ }\textbf
  {\bibinfo {volume} {104}},\ \bibinfo {pages} {085504} (\bibinfo {year}
  {2010})}\BibitemShut {NoStop}%
\bibitem [{\citenamefont {Wyart}\ \emph {et~al.}(2005)\citenamefont {Wyart},
  \citenamefont {Nagel},\ and\ \citenamefont {Witten}}]{wyart2005geometric}%
  \BibitemOpen
  \bibfield  {author} {\bibinfo {author} {\bibfnamefont {M.}~\bibnamefont
  {Wyart}}, \bibinfo {author} {\bibfnamefont {S.}~\bibnamefont {Nagel}}, \ and\
  \bibinfo {author} {\bibfnamefont {T.}~\bibnamefont {Witten}},\ }\href@noop {}
  {\bibfield  {journal} {\bibinfo  {journal} {Europhys. Lett.}\ }\textbf
  {\bibinfo {volume} {72}},\ \bibinfo {pages} {486} (\bibinfo {year}
  {2005})}\BibitemShut {NoStop}%
\bibitem [{\citenamefont {Lerner}\ \emph {et~al.}(2012)\citenamefont {Lerner},
  \citenamefont {D{\"u}ring},\ and\ \citenamefont {Wyart}}]{lerner2012unified}%
  \BibitemOpen
  \bibfield  {author} {\bibinfo {author} {\bibfnamefont {E.}~\bibnamefont
  {Lerner}}, \bibinfo {author} {\bibfnamefont {G.}~\bibnamefont {D{\"u}ring}},
  \ and\ \bibinfo {author} {\bibfnamefont {M.}~\bibnamefont {Wyart}},\
  }\href@noop {} {\bibfield  {journal} {\bibinfo  {journal} {Proc. Natl. Acad.
  Sci. U.S.A.}\ }\textbf {\bibinfo {volume} {109}},\ \bibinfo {pages} {4798}
  (\bibinfo {year} {2012})}\BibitemShut {NoStop}%
\end{thebibliography}%

\end{document}